\documentclass[12pt]{iopart}

\usepackage{iopams}
\usepackage{graphics}
\usepackage{epsfig}
\usepackage{amstext}

\newcommand{\ua}{\uparrow}
\newcommand{\da}{\downarrow}

\newcommand{\be}{\begin{equation}}
\newcommand{\ee}{\end{equation}}
\newcommand{\bea}{\begin{eqnarray}}
\newcommand{\eea}{\end{eqnarray}}

\newcommand{\vk}{{\boldsymbol k}}

\usepackage{color}
\usepackage{ulem} % riscar texto
\definecolor{green}{rgb}{0,0.75,0.3}

\begin{document}
\title{Magnetic chains on a triplet superconductor}
\author{ P. D. Sacramento }
%\email{ pdss@cfif.ist.utl.pt }
\address{ CeFEMA,
Instituto Superior T\'ecnico, Universidade de Lisboa, Av. Rovisco Pais, 1049-001 Lisboa, Portugal and}
\address{ Department of Physics, Kyoto University, Kyoto 606-8502, Japan
}
\ead{pdss@cfif.ist.utl.pt}
\date{ \today }

%%%%%%%%%%%%%%%%%%%%       ABSTRACT      %%%%%%%%%%%%%%%%%%%%

\begin{abstract}
The topological state of a two-dimensional triplet superconductor may be changed by an appropriate 
addition of magnetic impurities. A ferromagnetic magnetic chain at the surface of a superconductor
with spin-orbit coupling may eliminate
the edge states of a finite system giving rise to localized zero modes at the edges of the chain.
The coexistence/competition between the two types of zero modes is considered.
The reduction of the system to an effective $1d$ system gives partial information on the topological
properties but the study of the two sets of zero modes requires a two-dimensional treatment.
Increasing the impurity density from a magnetic chain to magnetic islands leads to a finite Chern number.
At half-filling small concentrations are enough to induce chiral modes.
\end{abstract}


\maketitle

\section{Introduction and model}

Topological systems are the focus of current great interest. Topological
superconductors \cite{Hasan,Zhang} have been studied in great detail, with particular emphasis on the
existence of Majorana fermions at its edges \cite{Alicea}, or located around local perturbations,
such as in the interior of vortices. A standard proposal is the one-dimensional
Kitaev model of spinless fermions with p-wave pairing \cite{Kitaev}, which displays
topological phase in some
parameter regimes, with the existence of localized zero energy
modes, if the chain is finite (and long enough). Several two-dimensional superconductors
also have topological properties such as the so-called $p+ip$ pairing \cite{pip,ivanov}. 
Adding spin-orbit interaction and a magnetic field either through its Zeeman effect or due to the presence
of vortices, a great variety of topological phases are predicted. 

Even though there are various candidates for a triplet superconductor, the proposal \cite{fukane} of
a conventional, very abundant in nature, s-wave singlet superconductor in proximity to 
a semiconductor wire in the presence of a Zeeman field \cite{dassarma}, that also has topological
properties, considerably increased the activity on this field. Experimental verifications
of the edge states have proved difficult but some experiments seem to provide good
evidence for their existence, in particular a nanowire on top of a superconductor \cite{mourik} and,
more recently, a set of (classical) magnetic impurities on top of a conventional superconductor
with their spin orientations arranged in some helical way \cite{bernevig1}, or a ferromagnetic chain 
in the additional presence of spin-orbit coupling \cite{science}. Some controversy
has however been raised \cite{finck,aguado,peng,dumitrescu}, even though several predictions suggest
their existence through different signatures, such as Andreev reflection \cite{tanaka,ana}.

Since both a triplet superconductor and a magnetic chain induce topological states, we explore
here the combined effect of the two by considering a set of magnetic impurities on top of
a triplet superconductor. 

The model considered here has the same structure of that considered in \cite{science}, with the
difference that the underlying superconductor has spin triplet pairing. However, the
inclusion of a Rashba like spin-orbit coupling implies the possible coexistence with
spin-singlet pairing \cite{Sigrist}, since parity is no longer conserved. If it exists it is assumed to be smaller
and is neglected here.

The system without the magnetic chain is described by the Hamiltonian \cite{sato}
\begin{eqnarray}
\hat H = \frac 1 2\sum_\vk  \left( {\boldsymbol \psi}_{\vk}^\dagger ,{\boldsymbol \psi}_{-\vk}   \right)
\left(\begin{array}{cc}
\hat H_0(\vk) & \hat \Delta(\vk) \\
\hat \Delta^{\dagger}(\vk) & -\hat H_0^T(-\vk) \end{array}\right)
\left( \begin{array}{c}
 {\boldsymbol \psi}_{\vk} \\  {\boldsymbol \psi}_{-\vk}^\dagger  \end{array}
\right)
\label{bdg1}
\end{eqnarray}
where $\left( {\boldsymbol \psi}_{\vk}^{\dagger}, {\boldsymbol \psi}_{-\vk} \right) =
\left( \psi_{\vk\ua}^{\dagger}, \psi_{\vk\da}^\dagger ,\psi_{-\vk\ua}, \psi_{-\vk\da}   \right)$
and
\begin{equation}
\hat H_0=\epsilon_\vk\sigma_0  + \hat H_R\,.
\end{equation}
Here, $\epsilon_{\boldsymbol{k}}=-2 t (\cos k_x + \cos k_y )-\mu$
is the kinetic part, $t$ denotes the hopping parameter set in
the following as the energy scale ($t=1$),
$\mu$ is the chemical potential,
$\boldsymbol{k}$ is a wave vector in the $xy$ plane, and we have taken
the lattice constant to be unity. 
The Rashba spin-orbit term is written as
\begin{equation}
\hat H_R = \boldsymbol{s} \cdot \boldsymbol{\sigma} = \alpha
\left( \sin k_y \sigma_x - \sin k_x \sigma_y \right)\,,
\end{equation}
 where $\alpha$ is measured in the same units.
The matrices $\sigma_x,\sigma_y,\sigma_z$ are
the Pauli matrices acting on the spin sector, and $\sigma_0$ is the
$2\times 2$ identity.

The pairing matrix reads
\begin{equation}
\hat \Delta = i\left( {\boldsymbol d}\cdot {\boldsymbol\sigma} \right) \sigma_y =
 \left(\begin{array}{cc}
-d_x+i d_y & d_z+\Delta_s \\
d_z-\Delta_s & d_x +i d_y
\end{array}\right)\,.
\end{equation}

The spin $S$ magnetic impurities act like local magnetic fields and are distributed
along a chain at the center of the two-dimensional system, now considered
finite with dimensions $N_x \times N_y$. Their contribution to the Hamiltonian
is written as
\be
H_m=-JS \sum_j \psi^{\dagger}(r_j) \left( \tau_z \otimes \sigma_z   \right) \psi(r_j)
\ee
where $\tau_i$ are the Pauli
matrices ($i=x,y,z$) acting on particle-hole space. Here $r_j$ are the
locations of the $N \leq N_x$ magnetic impurities distributed along the $x$ direction as
$r_j=x e_x$.

\section{Reduction to magnetic chain}

Since the main focus is on the low energy states of the system, we expect that these states will
be zero energy modes, that will either appear localized at the edges of the magnetic chain
or along the edges of the two-dimensional system, if open boundary onditions are used.
As we will see, the choice of boundary conditions naturally affects the solutions found
and it is the purpose of this work to determine the coexistence or competition between
the low energy states. For the moment let us focus on the states associated more directly
with the magnetic chain. This can be done by focusing on the degrees of freedom
at the locations of the spin impurities. One possible way is to integrate the degrees of
freedom of the rest of the system and to obtain an effective Hamiltonian for the remaining
degrees of freedom. This can be achieved, for instance, in a path integral formalism and
integrating out the degrees of freedom surrounding the magnetic chain, or using a lattice
Green's function method. A simpler method can be carried out by restricting the solutions
of the BdG equations to the low-energy (subgap) states of the magnetic chain.
As shown before for helical chains \cite{pientka} and also very recently for ferromagnetic
chains plus spin-orbit coupling \cite{srecent,hui}, 
the effective one-dimensional Hamiltonian will be of the type of the Kitaev model and
naturally leads to zero energy edge states.

We start from the Bogoliubov-de Gennes equations for the wave functions
that in the momentum representation are written as
\be
\label{bdg2}
\left(\begin{array}{cc}
\hat H_0(\vk) & \hat \Delta(\vk) \\
\hat \Delta^{\dagger}(\vk) & -\hat H_0^T(-\vk) \end{array}\right)
\left(\begin{array}{c}
u_n\\
v_n
\end{array}\right)
= w_{\vk,n}
\left(\begin{array}{c}
u_n\\
v_n
\end{array}\right).
\ee
The 4-component spinor can be written as
\be
\left(\begin{array}{c}
u_n\\
v_n
\end{array}\right)=
\left(\begin{array}{c}
u_n(\boldsymbol{k},\uparrow) \\
u_n(\boldsymbol{k},\downarrow) \\
v_n(-\boldsymbol{k},\uparrow) \\
v_n(-\boldsymbol{k},\downarrow) \\
\end{array}\right) .\ee
The impurity term is then an additional term to the Hamiltonian and will
change the BdG equations. Introducing the Fourier components of the local operators
at the impurity sites, we can write the BdG equations, including the magnetic impurities,
as \cite{pientka}
\be
\left( w_{\boldsymbol{k}}-H_{\boldsymbol{k}} \right) 
\psi_{\boldsymbol{k}} = -J S \sum_j \left( \tau_z \otimes \sigma_z \right) \psi(r_j) e^{-i k r_j}.
\ee
The left hand side may be written as the inverse of the Green function
\be
G_k^{-1}(w_k) \psi_k = -J S \sum_j \left( \tau_z \otimes \sigma_z \right) \psi(r_j) e^{-i k r_j}.
\ee
This leads to 
\be
\psi_k=-J S \sum_j G_k(w_k)  \left( \tau_z \otimes \sigma_z \right) \psi(r_j) e^{-i k r_j}.
\ee
Taking the inverse Fourier transform we may get that the wave functions for the impurity
locations are given by
\be
\psi(r_i) = -J S \sum_j \frac{1}{\sqrt{N}} \sum_k e^{ik(r_i-r_j)} G_k(w_k) 
\left( \tau_z \otimes \sigma_z \right) \psi(r_j).
\ee
Separating the $j=i$ term from the right hand side we get that
\be
F \psi(r_i) = -\sum_{j \neq i} \tilde{J}(i,j) \psi(r_j)
\ee
where
\be
F=\hat{I} + J S \frac{1}{\sqrt{N}} \sum_k G_k(w_k) \left( \tau_z \otimes \sigma_z \right)
\ee
and
\be
\tilde{J}(i,j)=J S \frac{1}{\sqrt{N}} \sum_k e^{ik(r_i-r_j)} G_k(w_k) \left( \tau_z \otimes \sigma_z \right).
\ee

Instead of attempting a direct solution of these equations
we will follow previous treatments \cite{pientka} that are suitable to obtain a solution
for the low energy subgap states. Indeed, since the interest is in
identifying the zero energy modes, a low energy expansion will be carried out.
Also, we consider the limit of weak coupling to the impurities. This implies
that in the term that connects the impurities, $\tilde{J}(i,j)$, we take the
energy to be zero, $\epsilon_k=0$, and in the term $F$ we take the limit of
small energies, considering only a linear term in the energy. In this way we may get 
an effective low energy eigenvalue equation where the energy is the eigenvalue
and the wave functions at the impurity locations are the eigenfunctions.
The point now is to identify the effective Hamiltonian for the magnetic
chain that replaces the full 2d original problem.

In general, and recalling that the Green function is for the system without
the magnetic impurities, we see that
\bea
G_k^{-1}(w) &=&  w \left( \tau_0 \otimes \sigma_0 \right)-\epsilon_k \left( \tau_z \otimes \sigma_0 \right)
-\alpha s_y \left( \tau_0 \otimes \sigma_x \right) -i \alpha s_x \left( \tau_z \otimes i \sigma_i \right)
\nonumber \\ 
&+& d_x \left( \tau_x \otimes \sigma_z \right) -i d_y \left( i \tau_y \otimes \sigma_0 \right)
-\Delta_s \left( i \tau_y \otimes i \sigma_y \right)
\eea
Here $s_x=\sin(k_x)$, $s_y=\sin(k_y)$ and we are considering systems with no $d_z$ component.
Let us now consider the spin singlet and the spin triplet cases separately.

In the case of the spin singlet ($d_x=d_y=0, \Delta_s \neq 0$) 
we get that the inverse of the Green's function is given by
(in the limit of small energies, $w$)
\bea
G  
= \frac{1}{D} \left(\begin{array}{cccc}
\epsilon_-  & i \alpha s_- z(-1,-1) & -2 i \alpha \Delta_s \epsilon_k s_-
& -\Delta_s z(1,1) \\
-i \alpha s_+ z(-1,-1) & \epsilon_-  & \Delta_s z(1,1) & -i 2\alpha \Delta_s \epsilon_k s_+ \\
2i\alpha \Delta_s \epsilon_k s_+ & \Delta_s z(1,1) & \epsilon_+  &
-i \alpha s_+ z(-1,-1) \\
-\Delta_s z(1,1) & i 2 \alpha \Delta_s \epsilon_k s_- & i\alpha s_- z(-1,-1) &
\epsilon_+ 
\end{array}\right) \nonumber \\
\eea
where $\epsilon_{\pm}=\pm \epsilon_k z(1,-1)-w z(1,1)$ with
$z(\eta,\eta^{\prime})=\epsilon_k^2+\eta \Delta_s^2 +\eta^{\prime} \alpha^2 (s_x^2+s_y^2)$,
$s_{\pm}=s_x \pm i s_y$ and
$D=z(1,1)^2-4\alpha^2\epsilon_k^2(s_x^2+s_y^2)$.
Note that the spin-orbit coupling ($\alpha \neq 0$) induces elements of
the spin triplet type (elements $(1,3),(2,4),(3,1),(4,2)$).

Multiplication by $\left( \tau_z \otimes \sigma_z \right)$ just changes the sign of all
elements of the second and third columns. 
In the approximation of low energy subgap states, the interaction term between the impurities
$\tilde{J}(i,j)$ does not depend on the energy, $w$, since we calculate the Green's function at zero energy.
We can rewrite the equation between the wave functions at the magnetic impurity locations as
\bea
& & \left(\begin{array}{ccccc}
\tilde{F} & \tilde{J}(1,2) & \tilde{J}(1,3) & \cdots & \tilde{J}(1,N) \\
\tilde{J}(2,1) & \tilde{F} &  \tilde{J}(2,3) & \cdots & \tilde{J}(2,N) \\
\tilde{J}(3,1) & \tilde{J}(3,2) & \tilde{F} &  \cdots & \tilde{J}(3,N) \\
\cdots & \cdots & \cdots & \cdots & \cdots \\
\tilde{J}(N,1) & \tilde{J}(N,2) & \tilde{J}(N,3) & \cdots & \tilde{F}
\end{array}\right)
\left(\begin{array}{c}
\psi(1) \\
\psi(2) \\
\psi(3) \\
\cdots \\
\psi(N)
\end{array}\right) \nonumber \\
&=& w J S \frac{1}{\sqrt{N}} \sum_k \frac{W(1,1)}{D}  \left( \tau_z \otimes \sigma_z \right)
\left(\begin{array}{c}
\psi(1) \\
\psi(2) \\
\psi(3) \\
\cdots \\
\psi(N)
\end{array}\right)
\eea
We see that the matrix on the left hand side acts like an effective Hamiltonian for the
magnetic chain wave functions. As shown before, the presence of the spin-orbit interaction
induces a pairing of the type $d_x,d_y$ and the s-wave singlet pairing gets renormalized, as
all the terms of the original Hamiltonian \cite{tanaka2}.

Considering now the case of triplet pairing and inserting the magnetic impurities, we have once
again to find the Green's function. This has now a simpler form, assuming the absence of the spin-orbit
interaction, since it is not needed to yield triplet pairing.
It is enough to consider the structure of the Green's function. This is given by
\be
G  = \frac{1}{D} \left(\begin{array}{cccc}
-\epsilon_k -w  & 0 & d_- & 0\\
0 & -\epsilon_k -w  & 0 & -d_+ \\
d_+ & 0 & \epsilon_k-w  & 0 \\
0 & -d_- & 0 & \epsilon_k -w
\end{array}\right)
\ee
where here $D=\epsilon_k^2 + d_x^2 + d_y^2 $ and $d_{\pm}=d_x \pm i d_y$. We see that the structure of the triplet pairing
is maintained and all terms get renormalized.

From these results we see that the two-dimensional problem has been approximately reduced to
a magnetic chain, with some effective Hamiltonian, that has the structure of a one-dimensional
triplet superconductor, and therefore similar to the Kitaev model. Therefore one expects some
regions in the phase diagram where the system may become topological and with the presence
of edge states, if the magnetic chain is finite. However, as noted before, there are more complex terms
with long range character both in the kinetic energy and in the pairing terms. These typically
decrease with distance but may have a relevant effect.

\section{Phase diagram of magnetic chain}

The effect of longer range interactions has been studied before in the
context of the Kitaev model \cite{niu,degottardi,wang} and in other topological superconducting systems
\cite{russo,luo}. The Kitaev model is mapped by a Jordan-Wigner transformation
to a spin problem, specifically the Ising model in a transverse field keeping
only nearest-neighbor couplings, but it maps to more general spin interactions
considering longer range terms. In these mappings the hopping and the pairing amplitude
are related, but one may consider a more general case where they can vary independently.
Taking, for instance, nearest neighbor and next-nearest neighbor terms (both in the
hopping term and in the pairing), leads to a rich phase diagram that has been studied
before. Interestingly, considering next-nearest neighbor terms leads to the possibility
of two Majorana fermions at each end of the chain, and including longer range terms
leads to an increasing number of Majoranas at the edges. In the limit where all first
terms are absent and a given higher order term is the first non-vanishing, the system
is equivalent to a set of copies of the original Kitaev chain (one for each
decoupled sublattice)  and therefore one has at each edge as many Majoranas as the number
of copies of the system.
The system of spinless fermions is in the BDI class with a $Z$ number of Majoranas at the edges
of the system if an effective time reversal symmetry holds and the states are protected due to
the existence of a chiral symmetry, which in this problem is just the particle-hole symmetry
of a BdG problem.

This has been determined before for the Kitaev model composed of spinless fermions.
The case of the magnetic chain is similar, but not quite the same. In this problem
we have spinful electrons and interactions that are more complex than just the triplet
pairing of the Kitaev model. We will determine the phase diagram of these magnetic chains
and also present some results, for the sake of completeness, for the Kitaev chain.
A chiral symmetry is also present enabling to classify the topological phases by a winding number.

A direct method to determine the regions in the phase diagram that are topological
and how many Majoranas are obtained at each edge, is to diagonalize the Hamiltonian
and determine its eigenstates and eigenvalues looking for zero modes. 
Another, more systematic way, is obtained calculating the
winding numbers that give directly the number of edge states in the problem \cite{tewari0,tewari}.
This is somewhat more reliable since the diagonalization is perfomed for a finite system
and, even though the energies of the Majorana states are rather low (in some cases of the
order of $10^{-16}$), in other cases they are higher (of the order of $10^{-5}$), 
but still the winding number is nonvanishing.

\begin{figure*}
\includegraphics[width=0.22\textwidth]{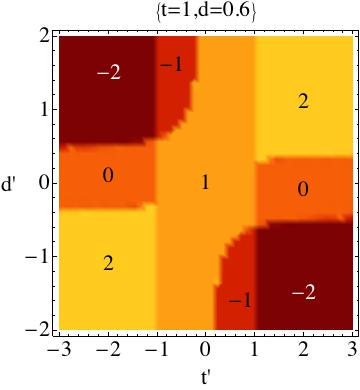}
\includegraphics[width=0.22\textwidth]{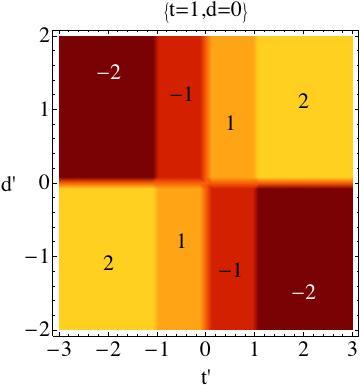}
\includegraphics[width=0.22\textwidth]{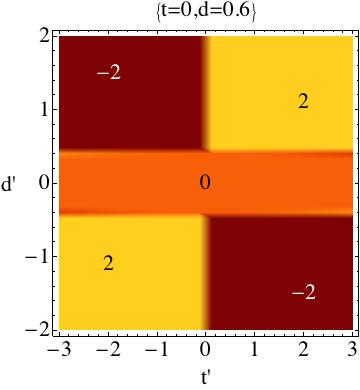}
\includegraphics[width=0.22\textwidth]{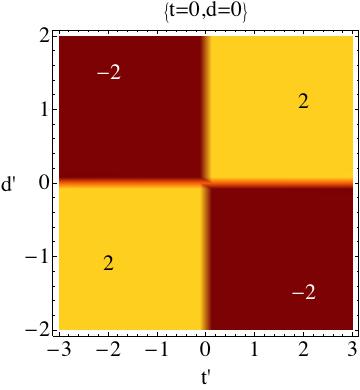}
$\mu=0$
\includegraphics[width=0.22\textwidth]{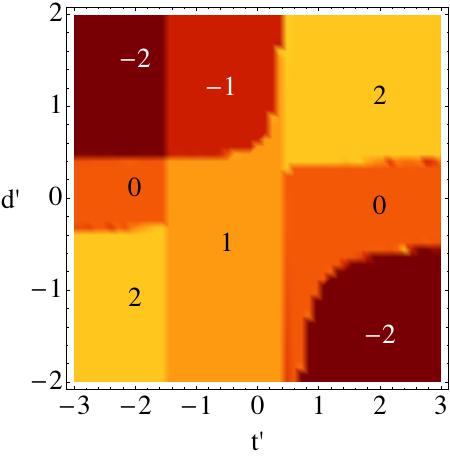}
\includegraphics[width=0.22\textwidth]{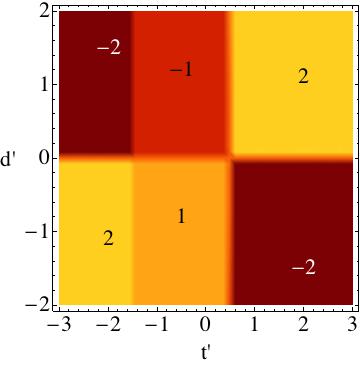}
\includegraphics[width=0.22\textwidth]{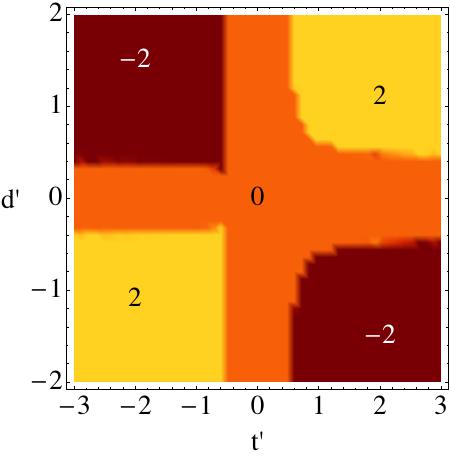}
\includegraphics[width=0.22\textwidth]{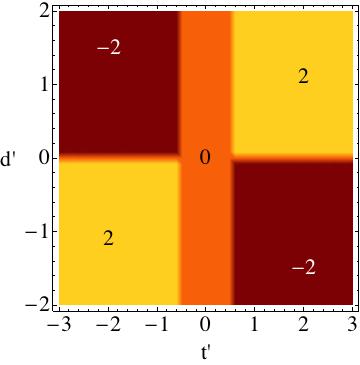}
$\mu=1$
\includegraphics[width=0.22\textwidth]{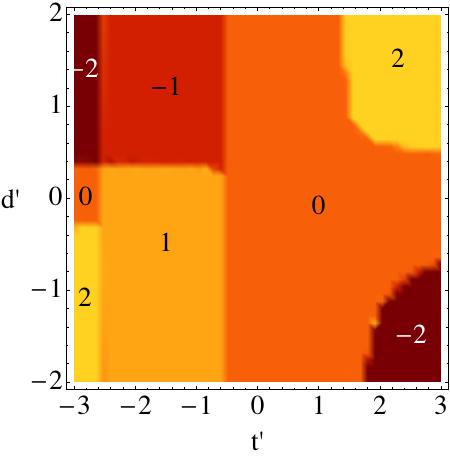}
\includegraphics[width=0.22\textwidth]{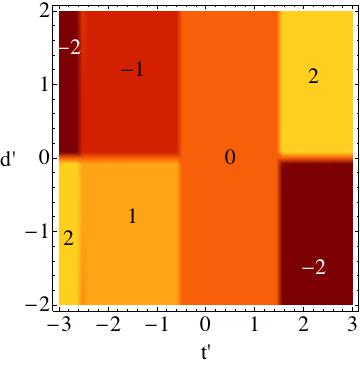}
\includegraphics[width=0.22\textwidth]{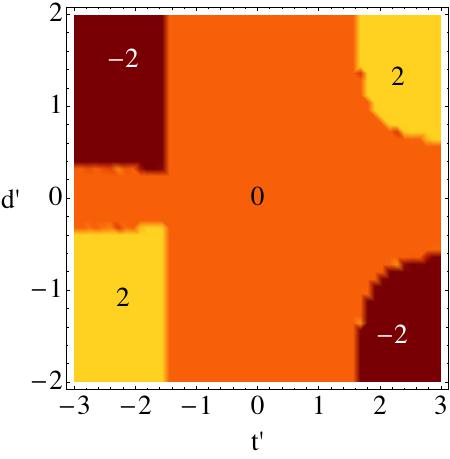}
\includegraphics[width=0.22\textwidth]{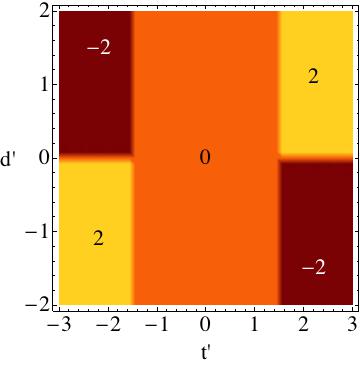}
$\mu=3$
\caption{\label{fig1}
(Color online) 
Phase diagrams for the extended Kitaev chain with second neighbor hopping and second neighbor
pairing, indexed by the winding number.
}
\end{figure*}

\subsection{1d extended Kitaev model}

We start with the case of the extended Kitaev model.
The model in real space may be written as
\bea
H=&-& t \sum_i \left( c_i^{\dagger} c_{i+1} + c_{i+1}^{\dagger} c_i \right) 
- t^{\prime} \sum_i \left( c_i^{\dagger} c_{i+2} + c_{i+2}^{\dagger} c_i \right) 
- \mu \sum_i c_i^{\dagger} c_i \nonumber \\
&+& \Delta \sum_i \left( c_i c_{i+1} + c_{i+1}^{\dagger} c_i^{\dagger} \right) 
+ \Delta^{\prime} \sum_i \left( c_i c_{i+2} + c_{i+2}^{\dagger} c_i^{\dagger} \right) 
\eea
where $t$ and $t^{\prime}$ are the nearest and next-nearest neighbor hoppings, respectively, and
$\Delta$ and $\Delta^{\prime}$ the nearest and next-nearest neighbor pairings, respectively.
This Hamiltonian may be diagonalized for a finite system of size $N_x$ using open boundary
conditions, and yields the presence of edge states of zero energy.

In momentum space the model is simply written as
\begin{eqnarray}
\hat H = \frac 1 2\sum_k  \left( c_k^\dagger ,c_{-k}   \right)
H_k
\left( \begin{array}{c}
c_{k} \\  c_{-k}^\dagger  \end{array}
\right)
\end{eqnarray}
where
\begin{eqnarray}
H_k = \left(\begin{array}{cc}
\epsilon_k -\mu & i \Delta \sin k+i\Delta^{\prime} \sin 2k \\
-i \Delta \sin k-i\Delta^{\prime} \sin 2k & -\epsilon_k +\mu  \end{array}\right)
\nonumber \\
\end{eqnarray}
with $\epsilon_k=-2t \cos k -2t^{\prime} cos 2k$.

Straightforward diagonalization gives
$\omega_k^2=\left(-\epsilon_k+\mu \right)^2+\left(\Delta \sin k +\Delta^{\prime} \sin 2k
\right)^2$.
Transitions between different topological phases occur when the system becomes gapless
which implies $-\epsilon_k+\mu=0$ and $\Delta \sin k +\Delta^{\prime} \sin 2k=0$.
This occurs when $\cos k=-\Delta/(2\Delta^{\prime})$, which leads to
$t^{\prime}=-\Delta (\mu \Delta^{\prime}-t \Delta)/(\Delta^2-2 (\Delta^{\prime})^2)$.
If $t^{\prime}=0,\Delta^{\prime}=0$, there are zeros for $k=0$ and $\mu=-2t$ or
$k=\pi$ and $\mu=2t$. If only $\Delta^{\prime}=0$ we get zeros at $k=0$ and
$2t+2t^{\prime}+\mu=0$ or $k=\pi$ and $-2t+2t^{\prime}+\mu=0$. If only $t^{\prime}=0$,
we get $\mu=t\Delta/\Delta^{\prime}$.

To calculate the winding number we follow the usual way by identifying a matrix that
anti-commutes with the Hamiltonian matrix. This is obtained verifying that
$\tau_x H_k \tau_x=-H$. Defining a matrix, $T$, composed by the eigenvectors of $\tau_x$, we
can reduce the Hamiltonian to an off-diagonal form
\begin{eqnarray}
T H T^{\dagger} = 
\left( \begin{array}{cc}
0 & q(k) \\  q^{\dagger}(k) & 0  \end{array}
\right)
\end{eqnarray}
where, in this case, $q(k)$ is just a function given by
\be
q(k)= \epsilon_k-\mu-i\Delta \sin k -i \Delta^{\prime} \sin 2k.
\ee
The winding number may be defined by
\be
I = \frac{1}{4\pi i} \int_{-\pi}^{\pi} Tr \left( q^{-1}(k) \frac{\partial}{\partial k}
q(k)-\left( q^{\dagger}\right)^{-1} \frac{\partial}{\partial k} q^{\dagger}(k) \right).
\ee

The phase diagram where the various topological phases are characterized by the 
winding number is represented in Fig. \ref{fig1}, where the results are presented
for various values of the chemical potential $\mu=0,1,3$. In the case
of only nearest-neighbor hopping and pairing, and $\mu=0,1$, we
are in topological phases and if $\mu=3$ we are in 
the topologically trivial phase (we take $t=1,d=0.6$).
Also, we consider cases in which one or both of
the nearest-neighbor terms are absent, namely $t=1,d=0$, $t=0,d=0.6$ and $t=0,d=0$.

The effective interaction $\tilde{J}(i,j)$ decays with distance and, as a result,
the nearest-neighbor terms are expected to be smaller than the nearest-neighbor ones
($t^{\prime}/t<1,\Delta^{\prime}/\Delta <1$). The results are however presented beyond this region 
for completeness. In the limits of high values of $t^{\prime}/t$ and/or $\Delta^{\prime}/\Delta$
one expects changes with respect to the original Kitaev model, as discussed above. However, the results
in Fig. \ref{fig1} show that, even in the regime where $t^{\prime}/t<1,\Delta^{\prime}/\Delta <1$, the
winding number changes with respect to the Kitaev model, and the topology and number of edge states changes.
For large values of $|t^{\prime}|$ or $|d^{\prime}|$, the winding number is $W=2,-2$ indicating two edge
states (on each end of the chain). Around the central region $W=1,0,-1$, as expected of the limit 
$t^{\prime}/t<<1,\Delta^{\prime}/\Delta <<1$. As shown here and shown before by other authors, 
the topology is changed in some cases even if the additional terms are small, 
rendering the system trivial or topological.
If $t=0$ the system is either trivial ($W=0$) or has two edge states ($W=2,-2$) for the various
chemical potential values. As the chemical potential grows the trivial phase becomes more widespread.

\begin{figure}
\includegraphics[width=0.4\columnwidth]{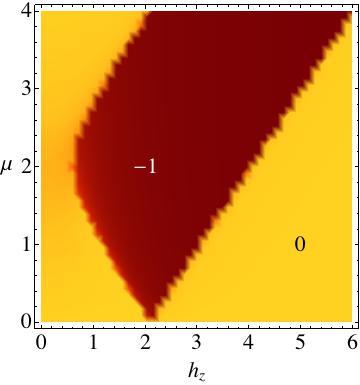}
\includegraphics[width=0.4\columnwidth]{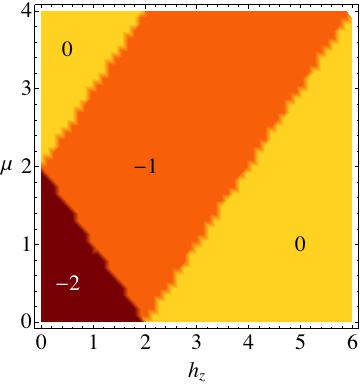}
\caption{\label{fig2}
(Color online) 
Phase diagram for an effective magnetic chain indexed by the winding number
for a $s$-wave superconductor (left panel) and a $p$-wave superconductor (right panel).
}
\end{figure}

\subsection{1d magnetic chain}

We consider now a magnetic chain described by an Hamiltonian matrix of the form 
\be
H_k  =  \left(\begin{array}{cccc}
\epsilon_k -h_z  & i\alpha \sin k & -id \sin k & \Delta_s \\
-i\alpha \sin k & \epsilon_k +h_z & -\Delta_s & -id \sin k \\
i d \sin k & -\Delta_s & -\epsilon_k +h_z  & -i\alpha \sin k \\
\Delta_s & id\sin k & i\alpha \sin k & -\epsilon_k -h_z
\end{array}\right)
\ee
where $\epsilon_k=-2t \cos k -\mu$ and, as above for the two-dimensional superconductor with magnetic
impurities, we consider the presence of spin-orbit coupling $\alpha$,
and triplet and singlet pairings along the direction of the chain;
$h_z$ is a local magnetic field that simulates the ferromagnetic ordering of
the impurities.
This model is just the one-dimensional reduction of the $2d$ Hamiltonian considered
in Eq. \ref{bdg1} (taking $k=k_x$).

The operator that anti-commutes with the Hamiltonian is now the matrix
$\tau_x \otimes \sigma_0$, and we obtain that $q(k)$ is a $2\times 2$ matrix
given by
\begin{eqnarray}
q(k) = \left(\begin{array}{cc}
\epsilon_k-h_z+id\sin k & i \alpha \sin k-\Delta_s \\
-i\alpha \sin k +\Delta_s & \epsilon_k +h_z+id\sin k  \end{array}\right)
\nonumber \\
\end{eqnarray}
where here $\epsilon_k=-2t\cos k -\mu$.
The winding number may be calculated in the same way.

The results are shown in Fig. \ref{fig2} as a function of $\mu$ and $h_z$, for a $s$-wave
superconductor and for a $p$-wave superconductor. 
In the case of the $s$-wave superconductor the winding number vanishes if the spin-orbit 
coupling is absent.
In the $p$-wave superconductor, even if the spin-orbit
coupling vanishes, the winding number is non-vanishing in regions of the phase
diagram.
If $\alpha \neq 0$ then the phase diagram does not depend on $\alpha$, for both
pairings. 

The phase diagram of the triplet chain is somewhat similar to the triplet $2d$ system \cite{sato} with a 
rescaling of both the chemical potential and the magnetic field from $4t$ to $2t$.
In the $2d$ case and $\mu>0$, there are two topologically distinct phases with $C=0$ and $C=-2$, where
$C$ is the Chern number, but both with two edge states. The same occurs in the magnetic chain.
For $\mu>2$ (and small $h_z$) and $h_z>2$ (small $\mu$), there are no edge states and the system is trivial.
There is also a region with $W=-1$ (one edge state) as for the $2d$ case.

The $s$-wave superconductor is however different from its $2d$ counterpart, since the phase
with two edge states ($W=\pm 2$) does not appear and only the phase with $W=-1$ is
topologically non-trivial.

\section{Two-dimensional system}

We consider now a finite two-dimensional system of dimensions $N_x \times N_y$, along a longitudinal direction
$x$, and a transversal direction $y$.
We apply different types of boundary conditions along the longitudinal and 
transverse directions leading to different sets of edge states.
We write
\be
\psi_{k_x,k_y,\sigma} = \frac{1}{\sqrt{N_y}} \sum_{j_y} e^{-i k_y j_y} \frac{1}{\sqrt{N_x}} \sum_{j_x} e^{-i k_x j_x}
 \psi_{j_x,j_y,\sigma}\,,
\label{operators2}
\ee
and rewrite the Hamiltonian matrix in terms of
the operators (\ref{operators2})  as
\bea
H = \sum_{j_x} \sum_{j_y}
& & \left(\begin{array}{cccc}
\psi_{j_x,j_y,\uparrow}^{\dagger}  & \psi_{j_x,j_y,\downarrow}^{\dagger} &
\psi_{j_x,j_y,\uparrow}  & \psi_{j_x,j_y,\downarrow}
\end{array}\right) \nonumber \\
& & \hat{H}_{j_x,j_y}
\left(\begin{array}{c}
\psi_{j_x,j_y,\uparrow} \\
\psi_{j_x,j_y,\downarrow} \\
\psi_{j_x,j_y,\uparrow}^{\dagger} \\
\psi_{j_x,j_y,\downarrow}^{\dagger} \\
\end{array}\right)
\eea

The operator $\hat{H}_{j_x,j_y}$ reads
\be
\hat{H}_{j_x,j_y}=
\left(\begin{array}{cc}
A & B \\
C & D \\
\end{array}\right)
\ee
where
\be
A=
\left(\begin{array}{cc}
-h_z-\epsilon_F-t \eta_+^x -t \eta_+^y & \frac{\alpha}{2} \eta_-^x +\frac{\alpha}{2i} \eta_-^y \\
-\frac{\alpha}{2} \eta_-^x +\frac{\alpha}{2i} \eta_-^y & h_z -\epsilon_F -t \eta_+^x -t \eta_+^y \\
\end{array}\right)
\ee
\be
B=
\left(\begin{array}{cc}
-\frac{d}{2} \eta_-^x -\frac{d}{2i} \eta_-^y & 0 \\
0 & -\frac{d}{2}\eta_-^x +\frac{d}{2i} \eta_-^y \\
\end{array}\right)
\ee
\be
C=
\left(\begin{array}{cc}
\frac{d}{2} \eta_-^x -\frac{d}{2i} \eta_-^y & 0 \\
0 & \frac{\alpha}{2} \eta_-^x +\frac{d}{2i} \eta_-^y \\
\end{array}\right)
\ee
\be
D=
\left(\begin{array}{cc}
h_z+\epsilon_F+t \eta_+^x + t \eta_+^y & -\frac{\alpha}{2} \eta_-^x +\frac{\alpha}{2i} \eta_-^y \\
\frac{\alpha}{2} \eta_-^x +\frac{\alpha}{2i} \eta_-^y & -h_z +\epsilon_F +t \eta_+^x +t \eta_+^y \\
\end{array}\right)
\ee
where $\psi_{j_x,j_y}^{\dagger} \eta_{\pm}^x \psi_{j_x,j_y} =
\psi_{j_x,j_y}^{\dagger} \psi_{j_x+1,j_y} \pm \psi_{j_x+1,j_y}^{\dagger} \psi_{j_x,j_y}$.
and $\psi_{j_x,j_y}^{\dagger} \eta_{\pm}^y \psi_{j_x,j_y} =
\psi_{j_x,j_y}^{\dagger} \psi_{j_x,j_y+1} \pm \psi_{j_x,j_y+1}^{\dagger} \psi_{j_x,j_y}$.
The diagonalization of this Hamiltonian involves the solution of a $(4 N_x N_y) \times (4 N_x N_y)$ eigenvalue problem.
The energy states include, in general, states in the bulk and states along the edges.
The local magnetic field describes different distributions of magnetic impurities, as needed.

\subsection{One impurity}

\begin{figure}
\includegraphics[width=0.95\columnwidth]{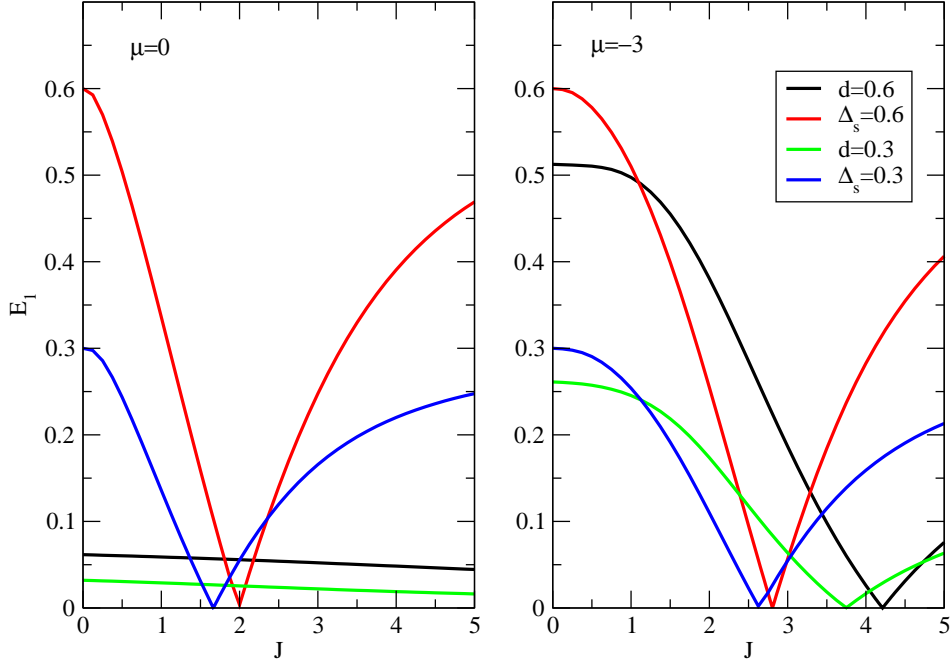}
\caption{\label{fig3}
(Color online) 
Lowest energy value as a function of the coupling between the impurity spin and the
spin density of the electrons.
}
\end{figure}

We begin by considering the effect of a single magnetic impurity on the $p$-wave superconductor and
compare the results with a magnetic impurity on a $s$-wave superconductor.

The effect of a single impurity in a conventional $s$-wave superconductor is well understood
\cite{imprmp}.
Since the impurity spin acts like a local magnetic field, the
electronic spin density will align along the local spin. For small
values of the coupling there is a negative spin density around the
impurity site. At the impurity site it is positive, as expected. For
larger couplings the spin density in the vicinity of the impurity
site is positive. At small couplings the many-body system screens
the effect induced by the impurity, inducing fluctuations that
compensate the effect of the local field, in a way that the overall
magnetization vanishes. However, for a strong enough coupling the
many-body system becomes magnetized in a discontinuous fashion. One
interpretation is that, if $J$ is strong enough, the impurity breaks a
Cooper pair and captures one of the electrons, leaving the other
electron unpaired, and thus the overall electronic system becomes
polarized. The impurity induces a pair of bound states inside the
superconducting energy gap, one at positive energy (with respect to
the chemical potential), and another at a symmetric negative energy.
Considering a higher value for the coupling one finds that
the levels inside the gap approach the Fermi level. There is a
critical value of the coupling for which the two levels cross in a
discontinuous way such that it coincides with the emergence of a
finite overall magnetization. After the level crossing occurs,
the nature of the states changes 
and, as the level crossing
takes place, the spin content also changes.
The level crossing occurs between one state that
describes an uncompensated local spin (at smaller coupling)
and a state where the impurity spin is compensated (partially since for the classical
description to be valid the spin has to be large).

In Fig. \ref{fig3} we show the results for the lowest (positive) energy state
for the $s$wave and the $p$-wave cases and for different chemical potentials
$\mu=0,-3$. In both cases, as a function of the coupling $J$, the lowest energy
state approaches zero and then increases again, signalling the quantum phase
transition (QPT). Note that here the calculation is not done self-consistently and
the superconductivity is assumed constant, as the result of a proximity effect
($\Delta_s,d=0.6$, respectively). In a full self-consistent solution the gap
function is decreased at the impurity site and at the quantum phase transition
changes sign ($\pi$ shift) and the energy level does not vanish in the $s$-wave
case. In the $p$-wave case the quantum critical point occurs at higher coupling
values and for $\mu=0$ it is rather flat up to high coupling values. 
Note that the effect of a single impurity is significant in the properties of the
system as shown before \cite{imprmp,mirage,ent1,fid1}, including its effect on transport 
properties inducing, for instance, currents \cite{annica}
and an anomalous Hall effect \cite{ahe}, if spin-orbit is present.

\begin{figure}
\includegraphics[width=0.45\columnwidth]{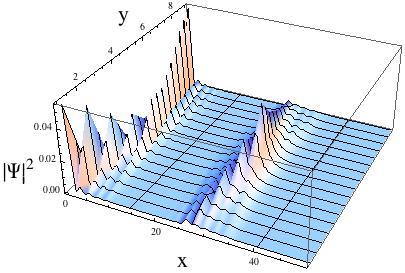}
\includegraphics[width=0.45\columnwidth]{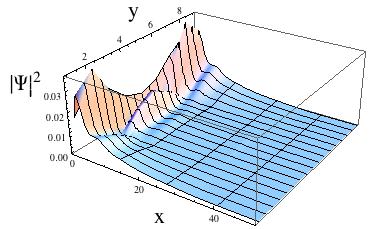}
\includegraphics[width=0.45\columnwidth]{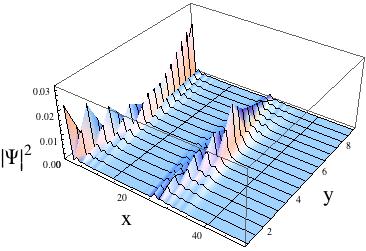}
\includegraphics[width=0.45\columnwidth]{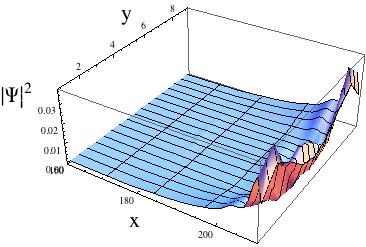}
\includegraphics[width=0.45\columnwidth]{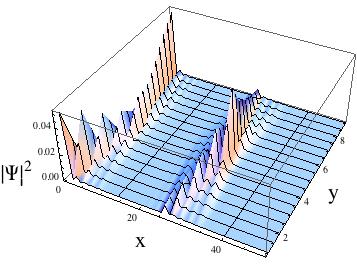}
\includegraphics[width=0.45\columnwidth]{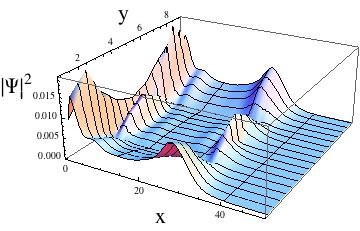}
\caption{\label{fig4}
(Color online) 
Lowest state wavefunctions for the case of OBC along $x$ and $y$ directions for $\mu=0$,
left column, and $\mu=-3$, right column, for $h_z=2,3,5$ for a system of size $210 \times 19$
and a chain of magnetic impurities of length $150$ sites ($N<N_x$). 
}
\end{figure}

\begin{figure}
\includegraphics[width=0.45\columnwidth]{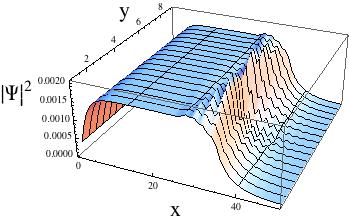}
\includegraphics[width=0.45\columnwidth]{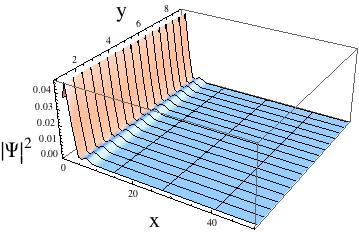}
\includegraphics[width=0.45\columnwidth]{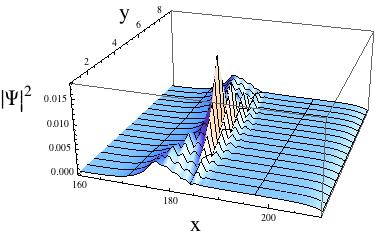}
\includegraphics[width=0.45\columnwidth]{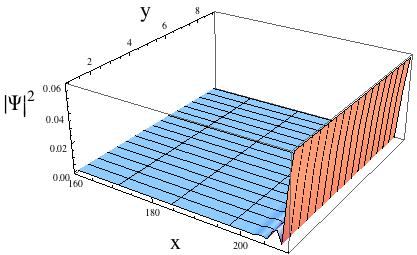}
\includegraphics[width=0.45\columnwidth]{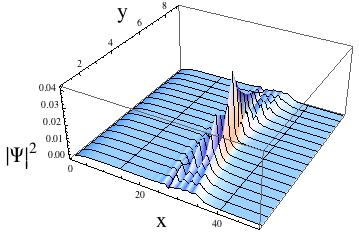}
\includegraphics[width=0.45\columnwidth]{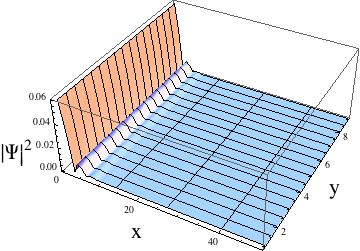}
\caption{\label{fig5}
(Color online) 
Lowest state wavefunctions for the case of OBC along $x$ and PBC along $y$, for $\mu=0$,
left column, and $\mu=-3$, right column, for $h_z=1,2,5$ for the same system of Fig. \ref{fig4}. 
}
\end{figure}

\subsection{Magnetic chains}

Increasing the number of impurities the number of in-gap states increases (two per
impurity) and the gap gets filled. If the impurities are arranged in a chain and
if this is long enough, Majorana edge states have been proposed to occur, if
for instance the alignment is parallel and there is spin-orbit coupling \cite{science}.

It is therefore interesting to see what happens if the magnetic impurities are
deposited on top of a $p$-wave superconductor. One expects two types of edge states.
Due to the magnetic chain one expects localized edge states at the ends of the
chain that, if the chain is long enough, are zero energy Majorana modes.
On the other hand, a $p$-wave superconductor at zero magnetic field has propagating
edge modes along the borders of a finite system in a stripe geometry (with periodic
boundary conditions along one direction and open boundary conditions along the other).
In order to observe the edge states along the chain we choose open boundary conditions
along the $x$ direction (magnetic chain direction) and along the $y$ direction we choose
either open boundary conditions or periodic boundary conditions. Also, we consider a case where
the magnetic chain is imbedded inside the two-dimensional system ($N<N_x$) and another case
where the magnetic chain extends up to the edges of the system ($N=N_x$). In the first case a spatial
separation between the two types of low energy edge states is possible while in the second
case, if both exist, they will be superimposed. In both cases there is a competition between
the two types of states, being more evident for the second case.

We consider first the case where the magnetic chain extends from site 30 up to site 150
along the $x$ direction, centered in the middle of the system in the $y$ direction, 
in a system of dimensions $210 \times 9$.
The lowest energy wave functions are either symmetric in $x$ or, if not, there is another
degenerate state or a corresponding state at negative energy that is its mirror. Since 
both types of edge states are rather localized, for better visualization we focus on a region
close to one border of the two-dimensional system.
Some states have very small energies but other edge states have subgap energies large but still
clearly smaller than the states in the continuum, typically at least one order of magnitude smaller.

It is perhaps helpful
to recall the phase diagram for a two-dimensional system with a uniform magnetic field \cite{sato,epl}.
In this work we have a non-homogeneous distribution of magnetic fields located at
the magnetic impurities, since these interact with the system as a local Zeeman term.
At $\mu=0,-3$ and zero magnetic field, the system is in a $Z_2$ phase with two degenerate edge modes  
on each edge, considering a stripe geometry. Turning on a magnetic field the system at $\mu=0$
changes to a $Z$ phase with Chern number $C=-2$, for any magnetic field between $0<h_z<4$. At $h_z=4$
there is a quantum phase transition to a trivial regime with $C=0$ and no edge states. In the case
$\mu=-3$ if $0<h_z<1$ the system has two edge states but $C=0$. In the region $1<h_z<3$ the system
has one edge state and $C=1$, between $3<h_z<7$ again one edge state but with $C=-1$, and above that
magnetic field the system becomes trivial with $C=0$ and no edge states.
In every change of the Chern number there is a quantum phase transition and the bulk gap closes.
The number of edge states may also be understood calculating the topological entanglement entropy \cite{tee}

\begin{figure}
\includegraphics[width=0.45\columnwidth]{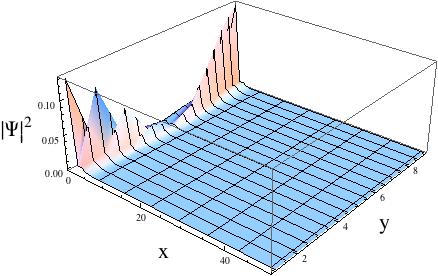}
\includegraphics[width=0.45\columnwidth]{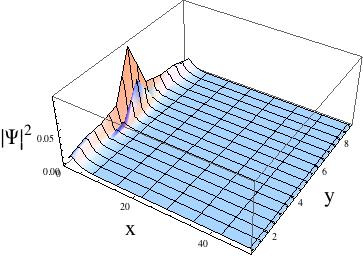}
\includegraphics[width=0.45\columnwidth]{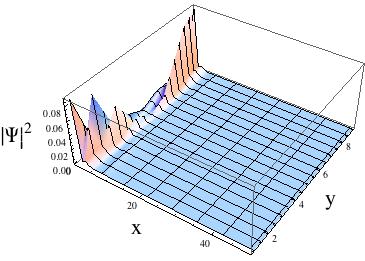}
\includegraphics[width=0.45\columnwidth]{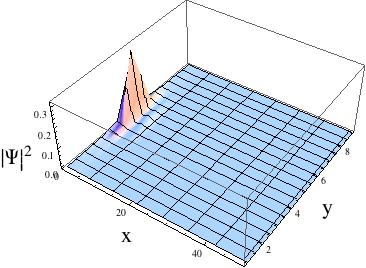}
\includegraphics[width=0.45\columnwidth]{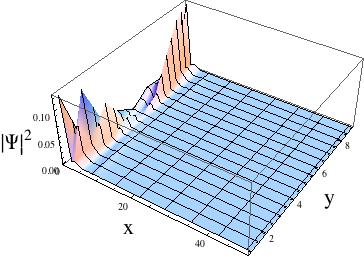}
\includegraphics[width=0.45\columnwidth]{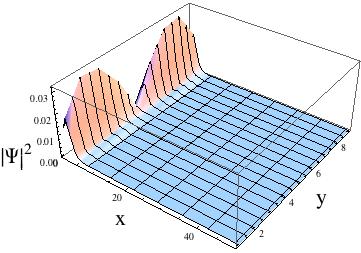}
\caption{\label{fig6}
(Color online) 
Lowest state wavefunctions for the case of OBC,OBC for $\mu=0$,
left column, and $\mu=-3$, right column, for $h_z=2,4,5$ for the edges of the magnetic chain
at the border of the system ($N=N_x$). 
}
\end{figure}

In Fig. \ref{fig4} we apply open boundary conditions (OBC) along both spatial directions
and show the density of the lowest state eigenfunction, as a function of
$x$ and $y$, for two values of the chemical potential $\mu=0,-3$, and consider that each
magnetic impurity can be seen as a local magnetic field with values $h_z=2,3,5$.
Also, we take $\alpha=0,d=0.6,\Delta_s=0$.
Consider first $\mu=0$. In the cases of $h_z=2,3,5$ the groundstate is four times degenerate, with an energy
of the order of $10^{-7}$ or $10^{-6}$, and the next set of states has an energy of the order of $10^{-2}$.
In the left column of Fig. \ref{fig4} one of the degenerate states is shown. In this case both types of
edge states are clearly seen. The state along the border of the system peaks at the edges along $y$ and the
state associated with the end of the magnetic chain is peaked at this edge and decays towards the edges along
the $y$ direction. Both states are very localized along the $x$ direction. 
If the chemical potential is $\mu=-3$, the border state dominates and any peak at the edge
of the magnetic chain is hardly visible at low magnetic fields, but its weight increases with the
magnetic field being clearly visible at $h_z=5$. 
However, in this case the two peaks overlap and in the
region between them the wave function is clearly finite and appreciable. 
The lowest energy state is non-degenerate and with very small energy of the order of $10^{-15}$
Actually, one may note that the two
peaks somewhat always overlap through the borders of the system along the $y$ direction
\cite{bernevig3}. Increasing the size along
$y$ does not diminish the spatial extent along $y$. 

Changing the boundary conditions along $y$ to periodic (PBC), as shown in Fig. \ref{fig5}, tends to separate
the two types of edge states.  In the case of $\mu=0$ the peak asociated with the magnetic chain
dominates (with now some overlap with the border state extending along $x$) and in the case of $\mu=-3$
the border states are even sharper. The border states are clearly invariant along $y$ due to the PBC but the
edge states at the chain ends are still peaked at the middle of the system. Note, however, that in the case of
$\mu=0$ the lowest energies are now quite high. The groundstate is doubly degenerate with an energy of the
order of $10^{-2}$. In the case of $\mu=-3$ the groundstate is now doubly degenerate with a rather small
energy of the order of $10^{-14}$ for $h_z=1,2$ and non-degenerate with an energy of the order of $10^{-18}$
at $h_z=5$.

Extending the magnetic chain up to the border of the system reverses the results between the two chemical
potentials: in the case of $\mu=0$ the border state dominates, while if $\mu=-3$ the chain edge 
state dominates,
as evidenced by the peak along $y$ centered at the chain location. These results are shown in Fig. \ref{fig6}.
Again for $\mu=0$ the energies of the lowest states are rather high, of the order of $10^{-2}$ for $h_z=2,4$, and
of the order of $10^{-3}$ for $h_z=5$, while for $\mu=-3$ they are nondegenerate for $h_z=2,4$ of the order of
$10^{-11}$ or smaller, while for $h_z=5$ the energy is of the order of $10^{-2}$ and the state is doubly
degenerate.

Associated with edge states we expect a bulk topological invariant. 
For translationally invariant systems a classification of the various possible symmetry
classes has been extensively studied \cite{schnyder1,schnyder2}.
In the presence of magnetic
field time reversal symmetry is broken and so a non-vanishing Chern number may be used
to classify the phases.
This has been used in the case of a uniform magnetic field, as reviewed above.
In the case of the magnetic chain immersed in the two-dimensional space, one may wonder
if the magnetic field applied on the system is enough to change its topology.
The usual way to calculate the Chern number, obtained through an integration over the
Brillouin zone, does not apply here since there is no translational invariance.
However, as discussed in the following subsection, one may define a Chern number in real
space by integrating over twisted boundary conditions. This method may be used for non-homogeneous
situations where crystal momentum is not a good quantum number.
A classification for systems with defects has also been carried out \cite{teokane}.

\subsection{Topological invariant: Chern number in real space}

A possible choice of a topological invariant to characterize each topological phase
is obtained calculating the Chern number associated with integration over twisted boundary conditions
in the two-dimensional system \cite{thouless}.

If a quantum system has an Hamiltonian that depends on some parameter $\boldsymbol R$ and if
it is periodic in this vector, than it can be shown that the integral of the $z$ component of the Berry
curvature over the surface the vector sweeps, $S$, is an integer called the Chern number
\be
C_n = \frac{1}{2 \pi} \int_S dS \Omega_z^{(n)} (\boldsymbol R ).
\ee
The index $n$ respects to a given eigenstate of the adiabatic energy levels
\be
H(\boldsymbol R ) |u_n(\boldsymbol R ) \rangle = E_n(\boldsymbol R ) |u_n(\boldsymbol R ) \rangle .
\ee
The z component of the Berry curbature of such a level is defined as (here $(1,2)=(x,y)$)
\be
\Omega_z^{(n)} = \frac{\partial}{\partial R_1} A_2^{(n)}(\boldsymbol R )-\frac{\partial}{\partial R_2} A_1^{(n)} (\boldsymbol R )
\ee
where the Berry connection is defined as
\be
\boldsymbol A^{(n)} (\boldsymbol R ) = \langle u_n(\boldsymbol R ) | i \nabla_{\boldsymbol R } | u_n(\boldsymbol R ) \rangle .
\ee
In a translationally invariant system the parameter $\boldsymbol R$ may be chosen as the momentum and the integration
runs over the Brillouin zone. In the problem at hand, which is not translationally invariant, the procedure may be
chosen in terms of twisted boundary conditions in both spatial directions. 

The integration over the Brillouin zone of the Berry curvature has in general numerical problems as one approaches
a transition point (for which the gap between bands closes) and it is more convenient to calculate the Chern
number in the lattice \cite{fukuis}.

The twisted boundary conditions are taken as
\be
u_n^{\theta}(\boldsymbol r +N_i \boldsymbol a_i) = e^{i \theta_i} u_n^{\theta} (\boldsymbol r )
\ee
where $\theta=(\theta_1,\theta_2)$ and $H(\theta)|u_n^{\theta} \rangle = E_n^{\theta} |u_n^{\theta} \rangle $, and $\boldsymbol a_i$ are the
basis vectors of the direct lattice, along direction $i$.
The problem is solved for a finite system with size $N_1 \times N_2$, typically taken large enough.
In general there are degeneracies in the energy spectrum which render the calculation of the Chern number difficult and so
a method on the lattice is also prefered. One considers the Slater determinant of the single-particle wave functions of
the Hamiltonian $H(\theta)$. The groundstate may then be represented by a matrix
\be
\Phi_{\theta}=
\left(\begin{array}{cccc}
\phi_{r_1}^{1,\theta} & \phi_{r_1}^{2,\theta} & \cdots & \phi_{r_1}^{M,\theta} \\
\phi_{r_2}^{1,\theta} & \phi_{r_2}^{2,\theta} & \cdots & \phi_{r_2}^{M,\theta} \\
\cdots & \cdots & \cdots & \cdots \\
\phi_{r_N}^{1,\theta} & \phi_{r_N}^{2,\theta} & \cdots & \phi_{r_N}^{M,\theta} \\
\end{array}\right)
\ee
where $N=N_1 N_2$ and $M$ is the number of points $M=M_1 M_2$ in the space of the twisted
boundary conditions. The lattice Chern number may then be obtained in a way similar
\be
C=\frac{1}{2\pi} \sum_{l=1}^{M_1 M_2} arg\left( 
\langle \Psi^{\theta_l} | \Psi^{\theta_l+\boldsymbol 1 } \rangle
\langle \Psi^{\theta_l+\boldsymbol 1 } | \Psi^{\theta_l+\boldsymbol 1 +\boldsymbol 2} \rangle
\langle \Psi^{\theta_l+\boldsymbol 1 +\boldsymbol 2} | \Psi^{\theta_l+\boldsymbol 2 } \rangle
\langle \Psi^{\theta_l+\boldsymbol 2 } | \Psi^{\theta_l} \rangle
\right)
\ee
where the vectors $\boldsymbol 1$ and $\boldsymbol 2 $ denote the two spatial directions corresponding
to the two types of twisted boundary conditions along the system's spatial directions.
Since each state is now a many-body state given by a Slater determinant, the overlaps between two
states with different boundary conditions are given by
\be
\langle \Psi^{\theta} | \Psi^{\theta^{\prime}} \rangle
=det \left( \Phi_{\theta}^{\dagger} \Phi_{\theta^{\prime}} \right)
\ee
and the Chern number is obtained by
\be
C = \frac{1}{2\pi} \sum_{l=1}^{M} arg \lambda_p
\ee
where $\lambda_p$ are the eigenvalues of the matrix product
\be
\prod_{l=1}^{M} \Phi_{\theta_l}^{\dagger} \Phi_{\theta_l +\boldsymbol 1}
 \Phi_{\theta_l+\boldsymbol 1}^{\dagger} \Phi_{\theta_l +\boldsymbol 1 +\boldsymbol 2}
 \Phi_{\theta_l+\boldsymbol 1 +\boldsymbol 2}^{\dagger} \Phi_{\theta_l +\boldsymbol 2}
 \Phi_{\theta_l+\boldsymbol 2}^{\dagger} \Phi_{\theta_l}
\ee
However, this procedure is very time consuming. It has been shown that it is enough to
calculate the Chern using only periodic boundary conditions and not twisted boundary
conditions \cite{china}.
The Chern number is obtained by a similar expression but the eigenvalues $\lambda_p$ are the eigenvalues
of the $M \times M$ matrix
\be
F=C_{q_0 q_1} C_{q_1 q_2} C_{q_2 q_3} C_{q_3 q_0}
\ee
where the momenta are defined by
\be
\boldsymbol q_0=(0,0); \boldsymbol q_1=(\frac{2\pi}{N_1},0); \boldsymbol q_2=(\frac{2\pi}{N_1},\frac{2\pi}{N_2});
\boldsymbol q_3=(0,\frac{2\pi}{N_2}).
\ee
The matrices are defined by its matrix elements, $m n$, as
\be
C_{\boldsymbol q, \boldsymbol q^{\prime}}^{m n} = \sum_{\boldsymbol r_i}
\left(\phi_{\boldsymbol r_i}^{m,\theta=0}\right)^* e^{i(\boldsymbol q - \boldsymbol q^{\prime}) \cdot \boldsymbol r_i}
\phi_{\boldsymbol r_i}^{n,\theta=0}.
\ee

\begin{figure}
\includegraphics[width=0.45\columnwidth]{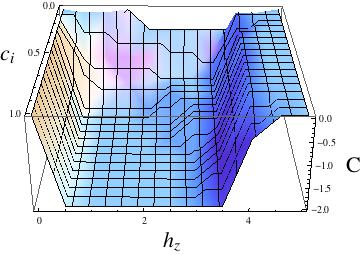}
\includegraphics[width=0.45\columnwidth]{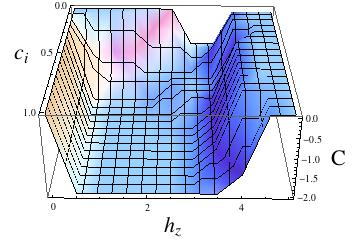}
\caption{\label{fig7}
(Color online) 
Chern number as a function of concentration and magnetic field for magnetic
islands of increasing size (concentration) for $\mu=0$ and $\alpha=0,0.6$ (left and right panels,
respectively). 
}
\end{figure}

\begin{figure}
\includegraphics[width=0.45\columnwidth]{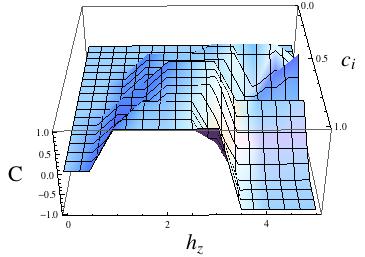}
\includegraphics[width=0.45\columnwidth]{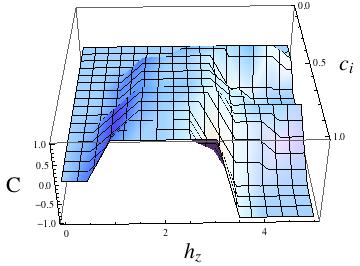}
\caption{\label{fig8}
(Color online) 
Chern number as a function of concentration and magnetic field for magnetic
islands of increasing size (concentration) for $\mu=-3$ and 
$\alpha=0,0.6$ (left and right panels, respectively). 
}
\end{figure}

Applying this method to the problem of the magnetic chain yields zero Chern number, unless the magnetic field
is very large, since the magnetic impurity concentration is small, and if we can think in a mean-field like
argument the average magnetic field is small. One expects therefore that in the case of $\mu=-3$ the Chern number
should vanish. In the case of $\mu=0$, since in a uniform magnetic field any small field will
change the Chern number from $C=0$ to $C=-2$ the Chern number may be finite. 
In order to better understand the effect of the impurity concentration,
we consider in the next subsection magnetic islands of growing size, centered in the $2d$ system. 
A system of various chains has been studied recently for a singlet superconductor \cite{kim}.
Also, edge states at the borders of a superconducting island in a topological insulator
\cite{nori} and a magnetic island on a singlet superconductor \cite{bernevigi} have been
studied very recently.

\subsection{Magnetic islands}

We consider squares of increasing size centered in the two-dimensional system
with dimensions $(2m+1) \times (2m+1)$, where $m=1,\cdots ,10$ in a system
of $21 \times 21$ sites.
The Chern number can then be calculated for these regular distributions of magnetic impurities.
The results are shown in Figs. \ref{fig7}, \ref{fig8} as a function of the magnetic field
due to each impurity and the concentration ($c_i=(2m+1)^2/21^2$).

\begin{figure*}
\includegraphics[width=0.32\textwidth]{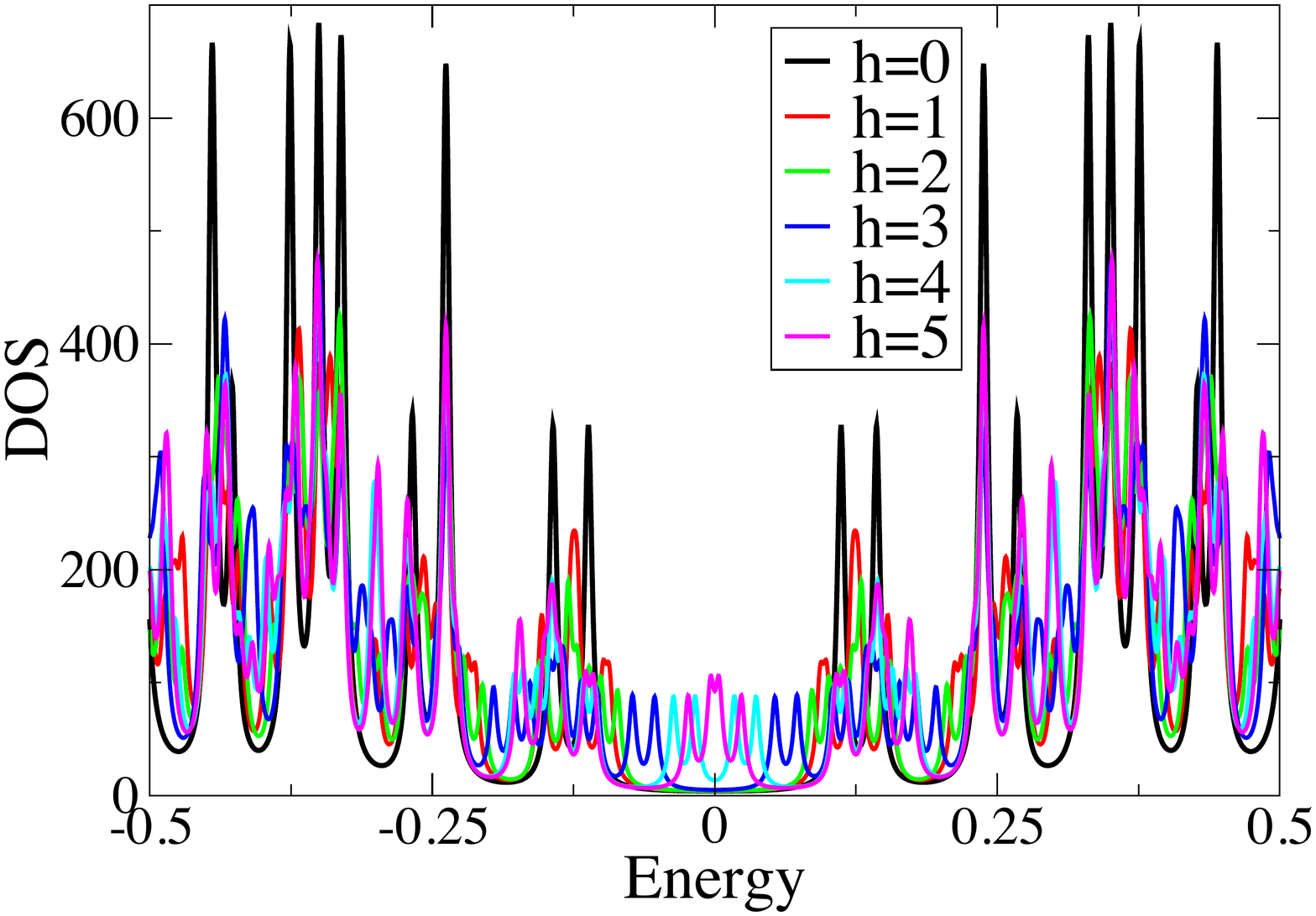}
\includegraphics[width=0.32\textwidth]{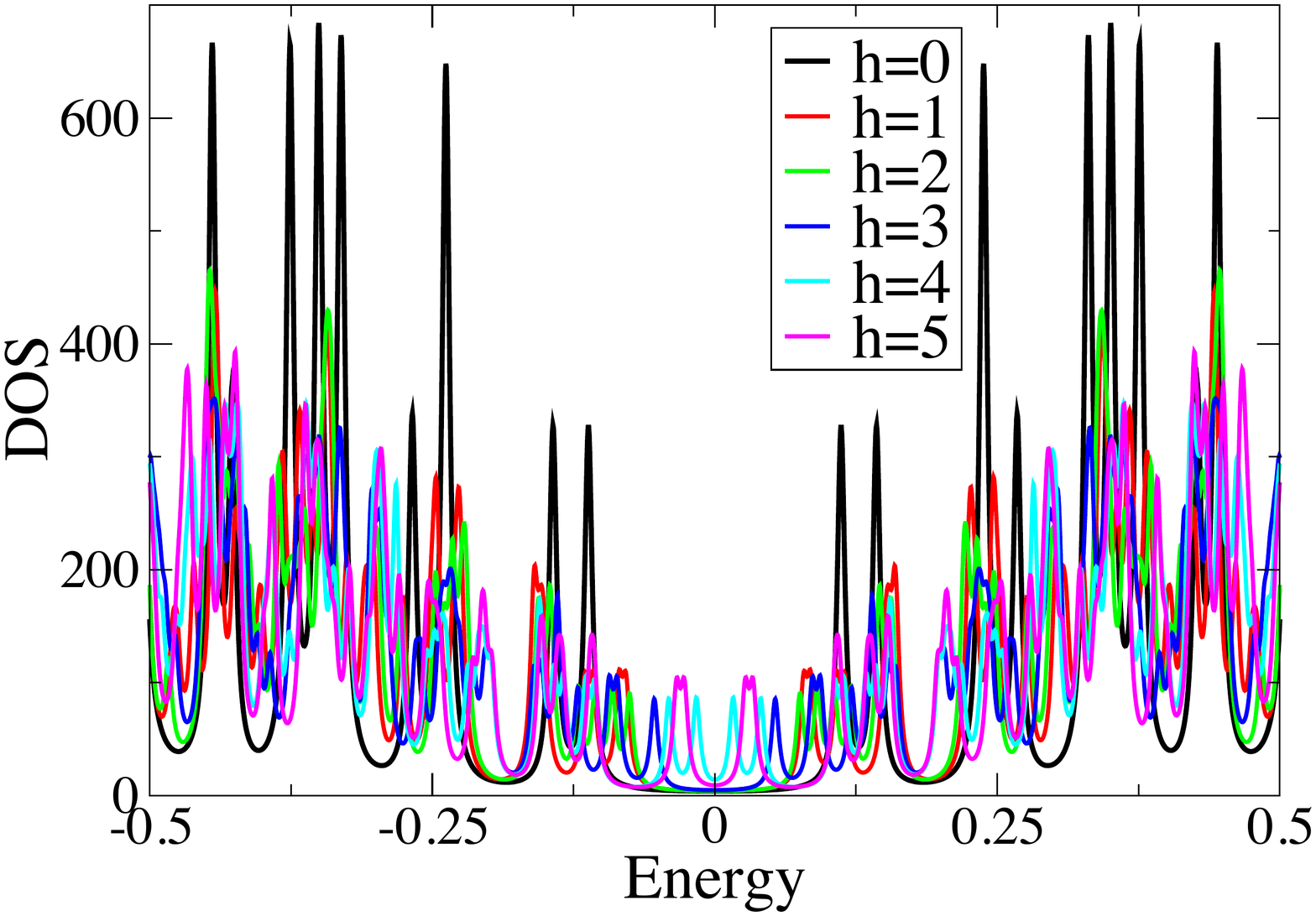}
\includegraphics[width=0.32\textwidth]{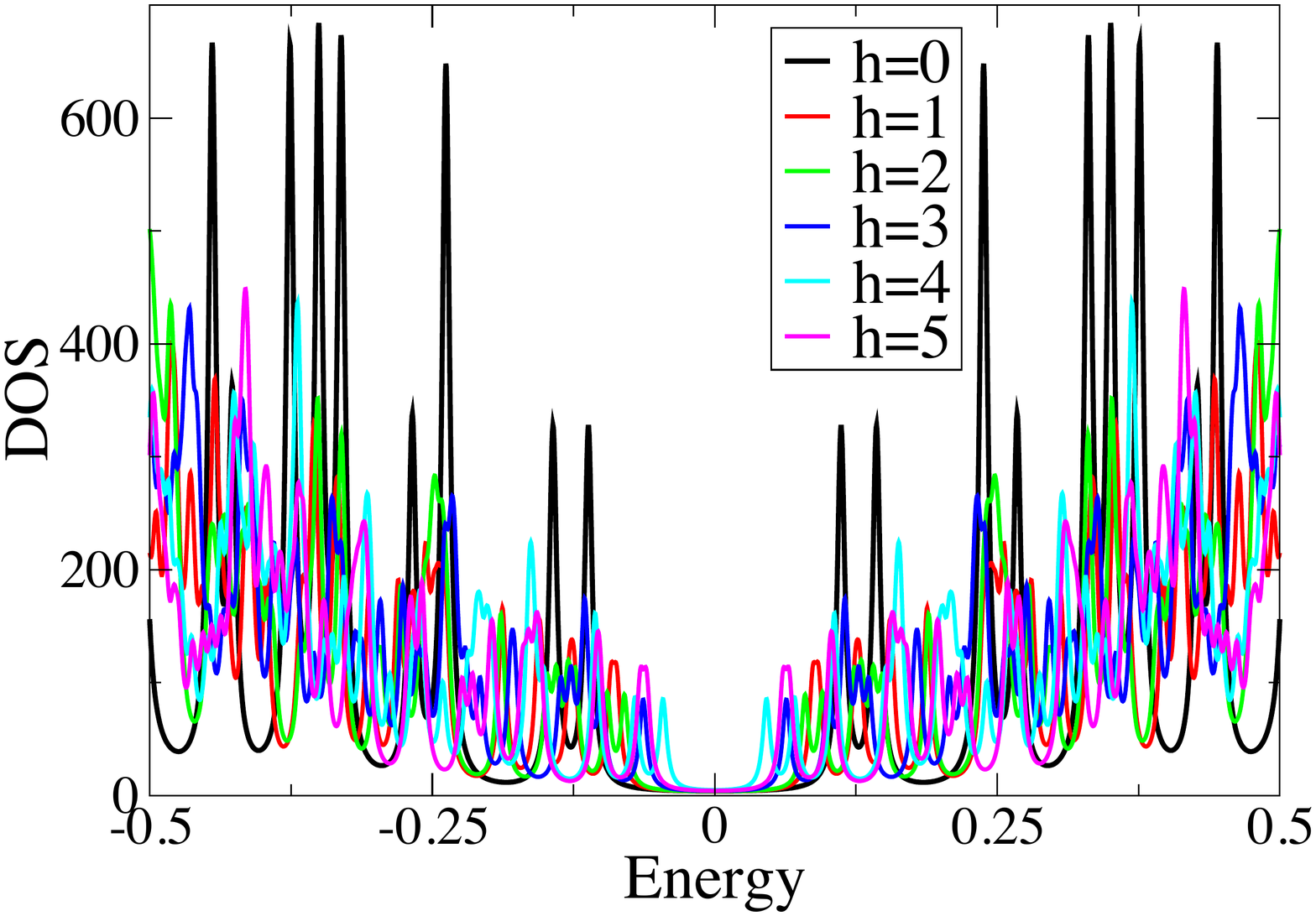}
\includegraphics[width=0.32\textwidth]{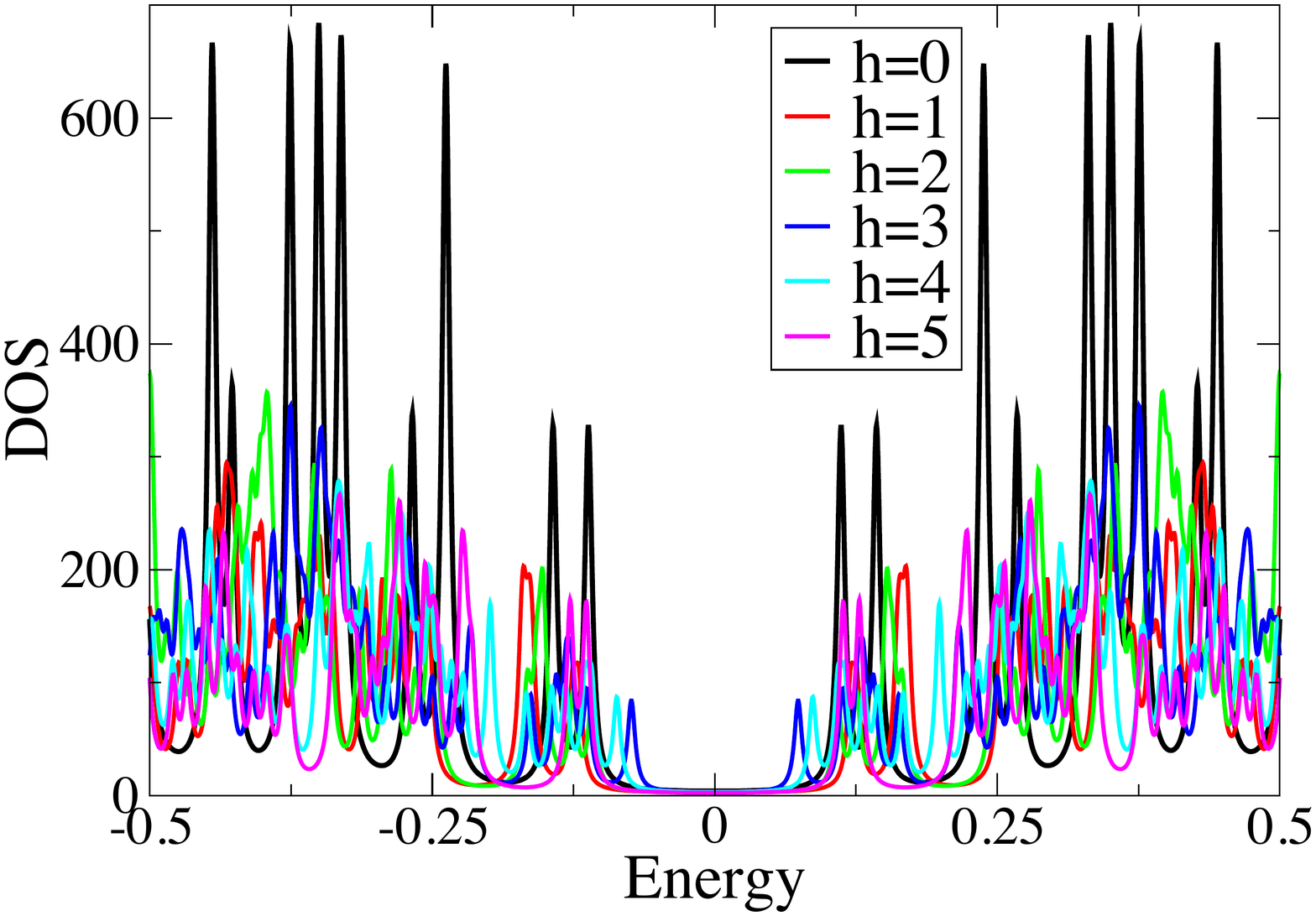}
\includegraphics[width=0.32\textwidth]{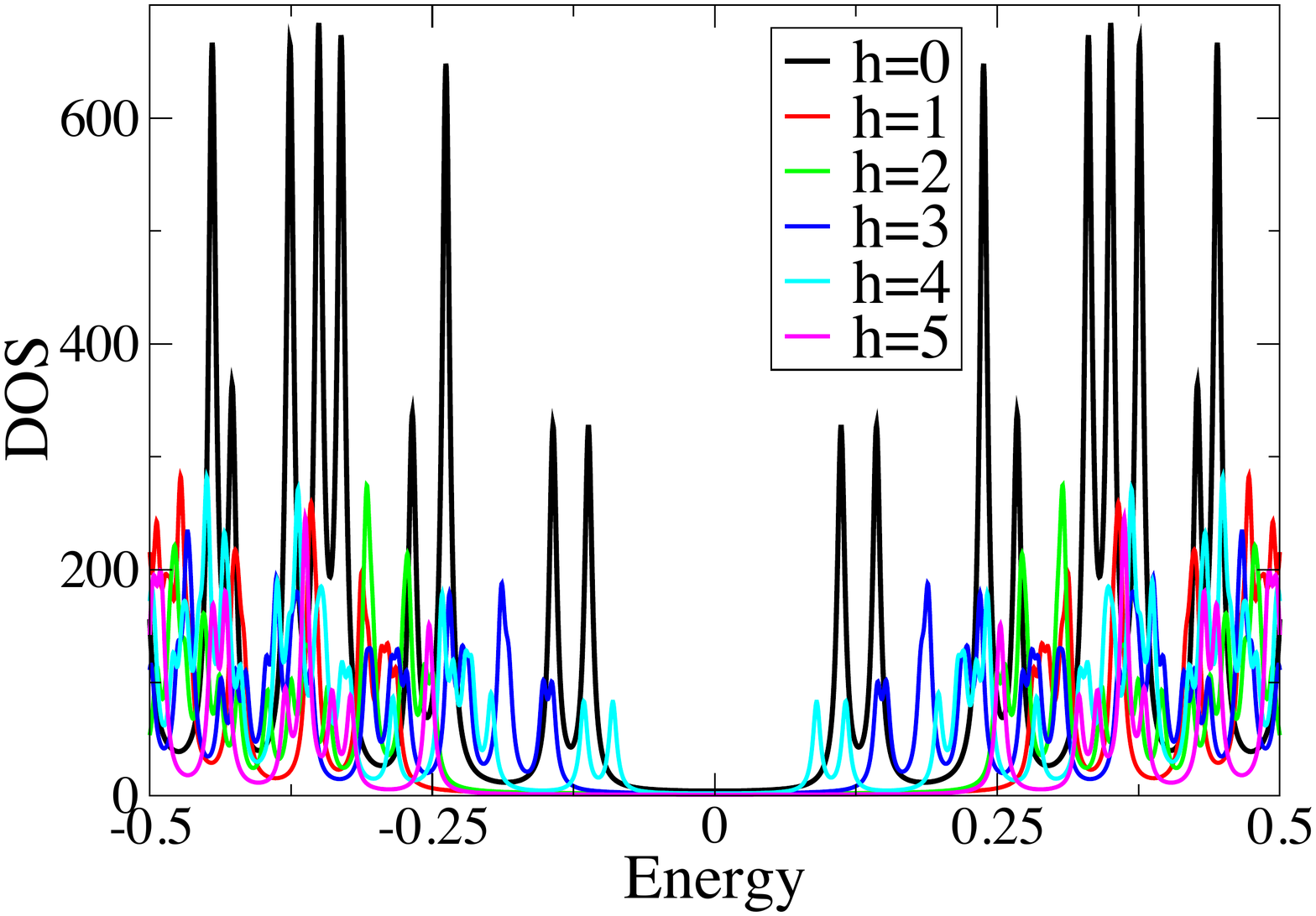}
\includegraphics[width=0.32\textwidth]{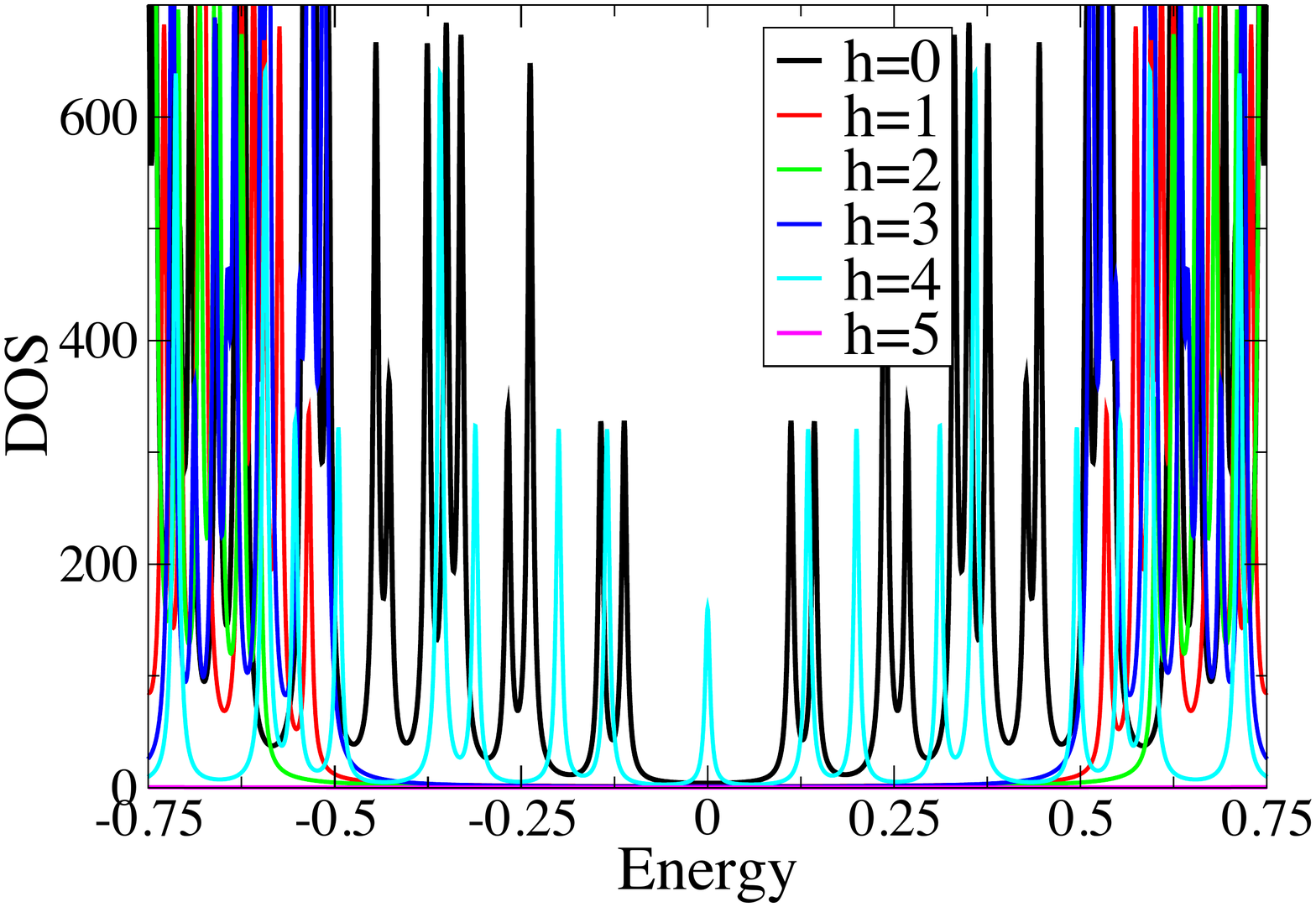}
\caption{\label{fig9}
(Color online) 
DOS for various concentrations $c_i=0.02,0.11,0.27,0.51,0.82,1$ for
$\mu=0$ and various magnetic fields $h_z=0,1,2,3,4,5$.
}
\end{figure*}

\begin{figure*}
\includegraphics[width=0.32\textwidth]{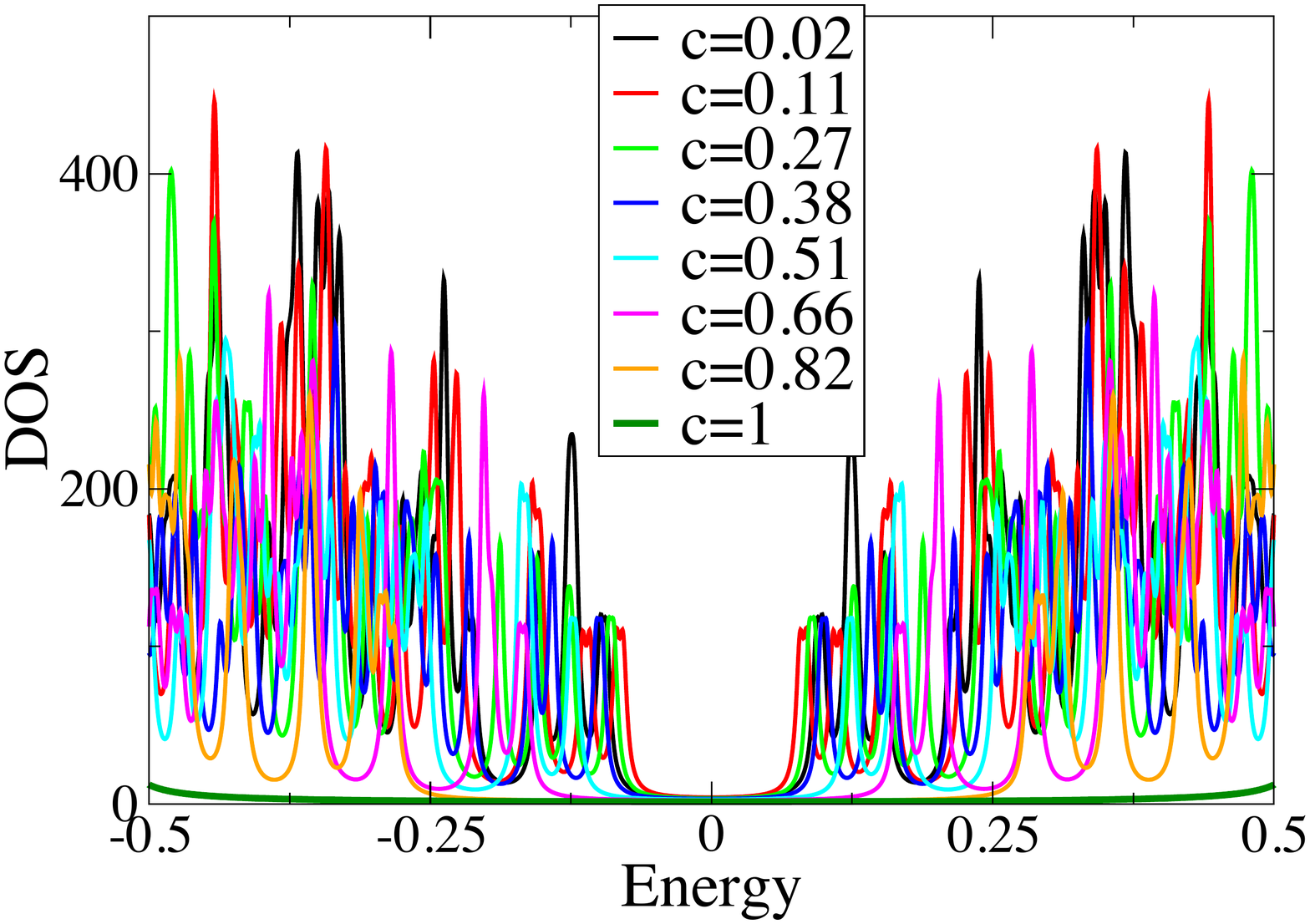}
\includegraphics[width=0.32\textwidth]{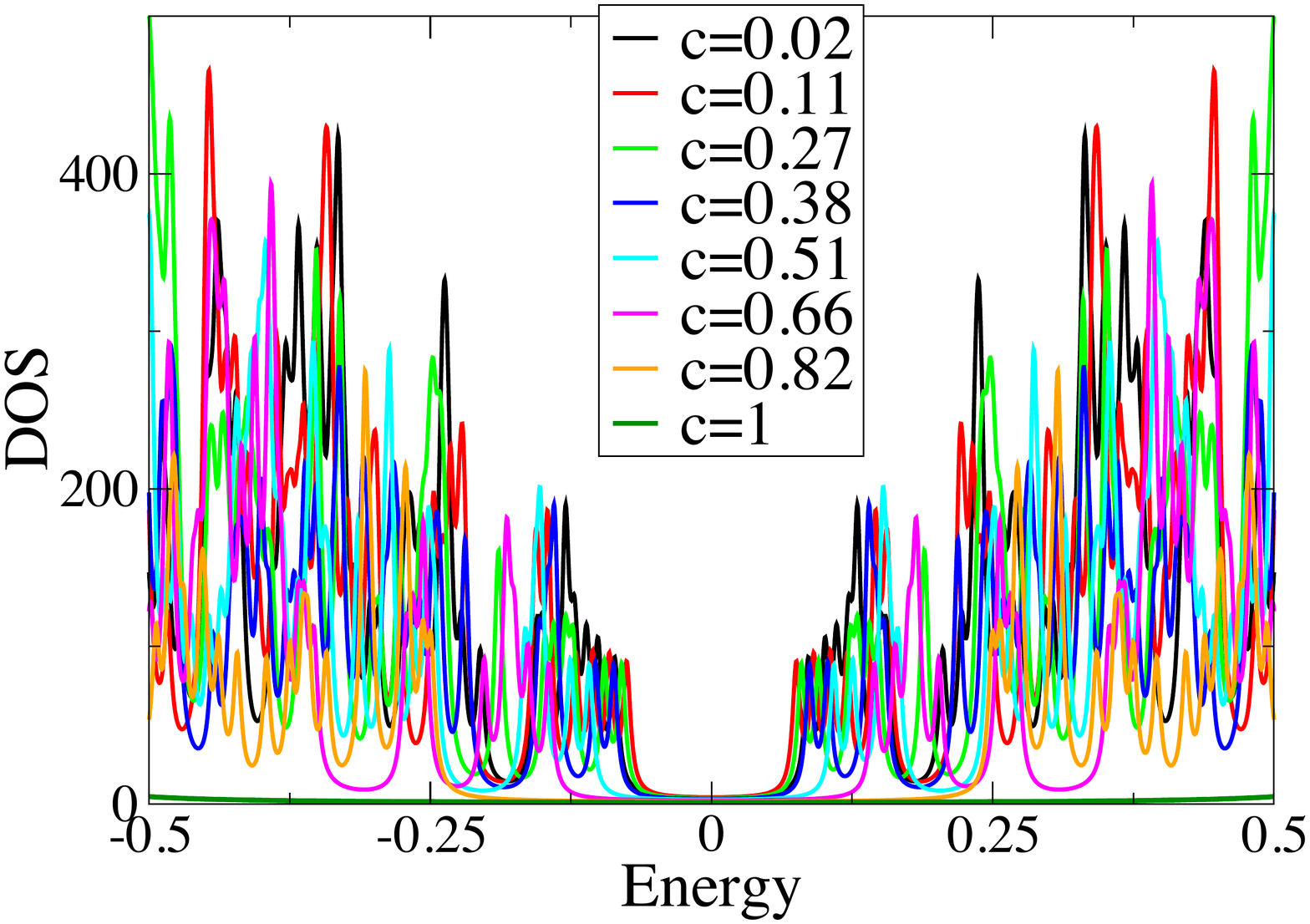}
\includegraphics[width=0.32\textwidth]{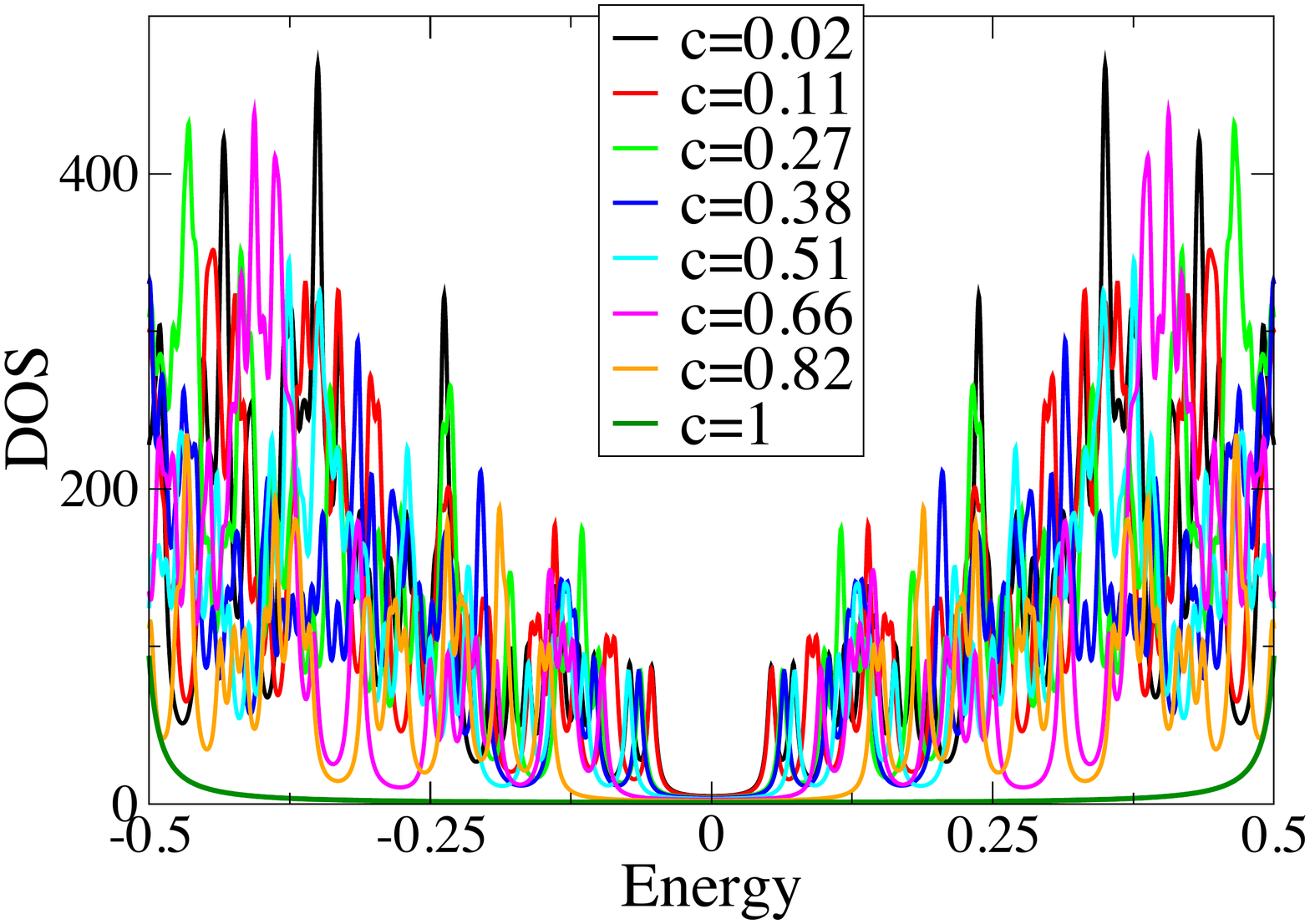}
\includegraphics[width=0.32\textwidth]{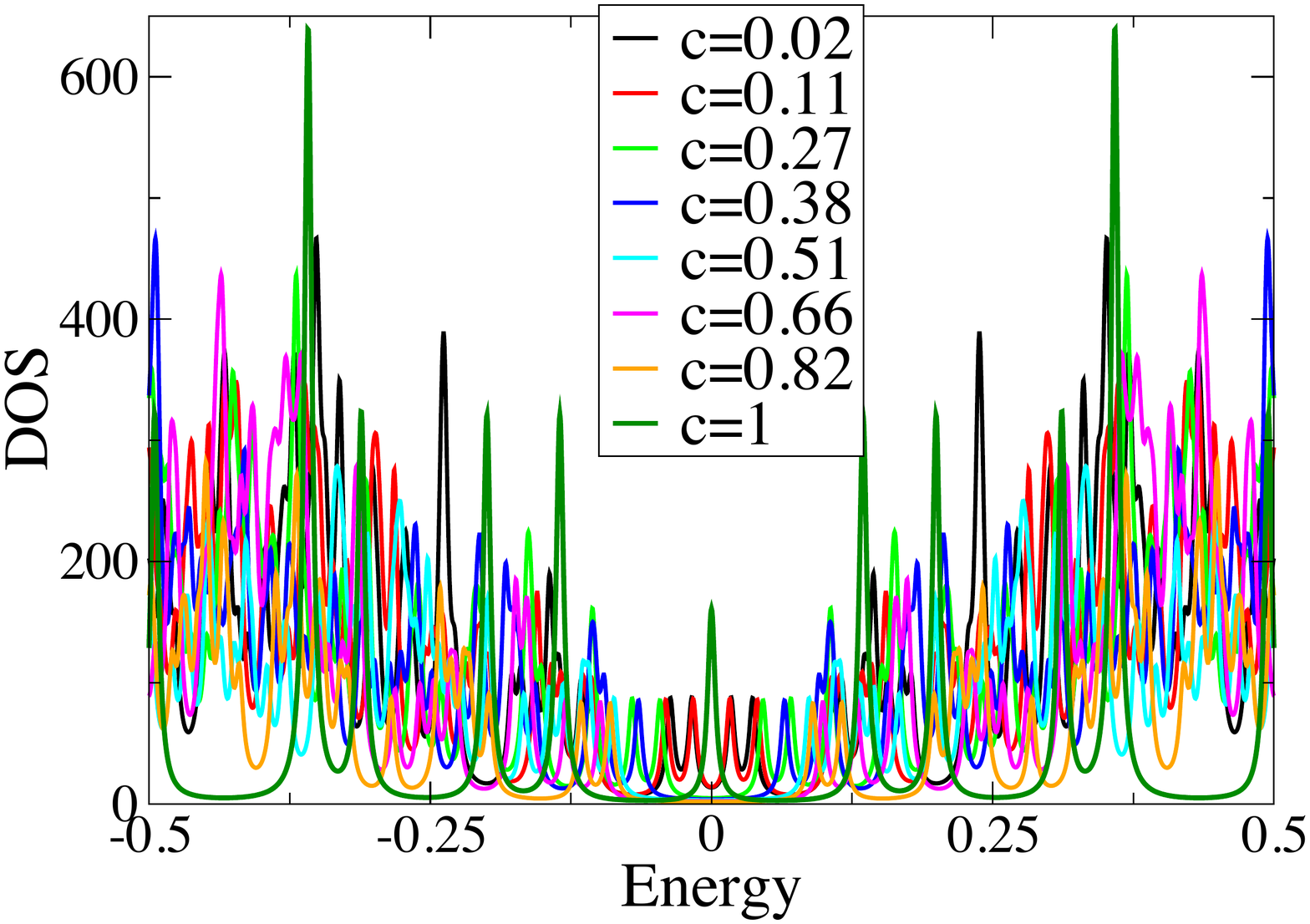}
\includegraphics[width=0.32\textwidth]{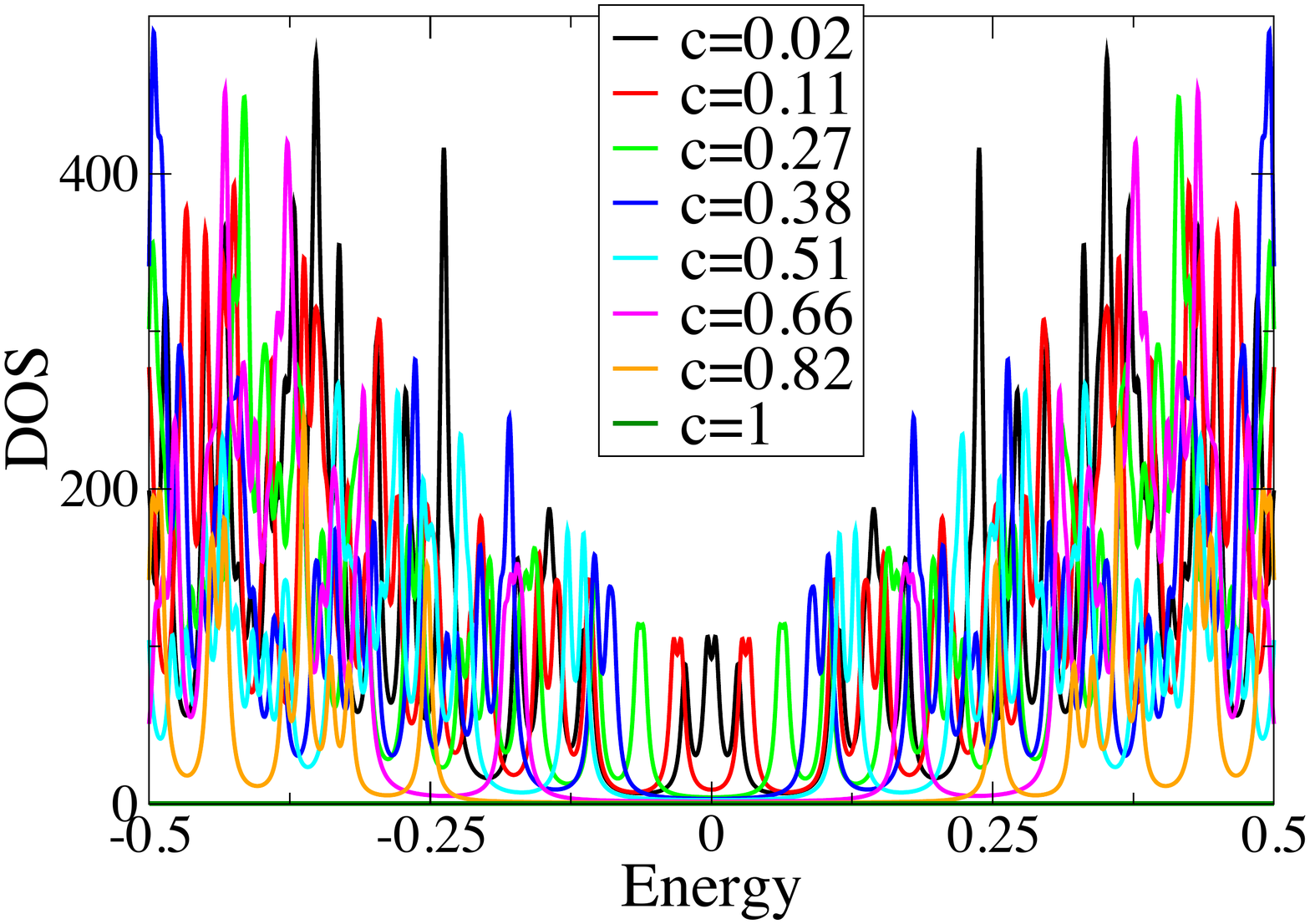}
\caption{\label{fig10}
(Color online) 
DOS for 
various magnetic fields $h_z=1,2,3,4,5$
and various concentrations 
$c_i=0.02,0.11,0.27,0.38,0.51,0.66,0.82,1$ for
$\mu=0$. 
}
\end{figure*}

The case of concentration $c_i=1$ is the case of a uniform magnetic field. In this case
the results using the real space representation of the Chern number, agree remarkably well
with the results obtained in momentum space. Lowering the concentration the same trend
is still observed. At zero magnetic field clearly $C=0$. At small concentrations the calculation
of the Chern is unstable and gives results that are between the results for $c_i=1$. The case of
$\mu=0$ is more stable, as expected. The unstable results for the Chern occur particularly
in the regions where the Chern is changing its value. However, the results for the Chern are often
half-integer indicating non-analyticies, as expected since there is a quantum phase transition.
In the uniform case this is a sharp transition with the closing of the bulk gap but, for the
non-homogeneous case, the transition regions are quite extended, particularly as the concentration
decreases, the magnetic field is large and particularly for $\mu=-3$ (as compared to $\mu=0$). 
These instabilities are associated
with states in the gap approaching zero, as will be seen in the calculations of the density of states.

\begin{figure*}
\includegraphics[width=0.32\textwidth]{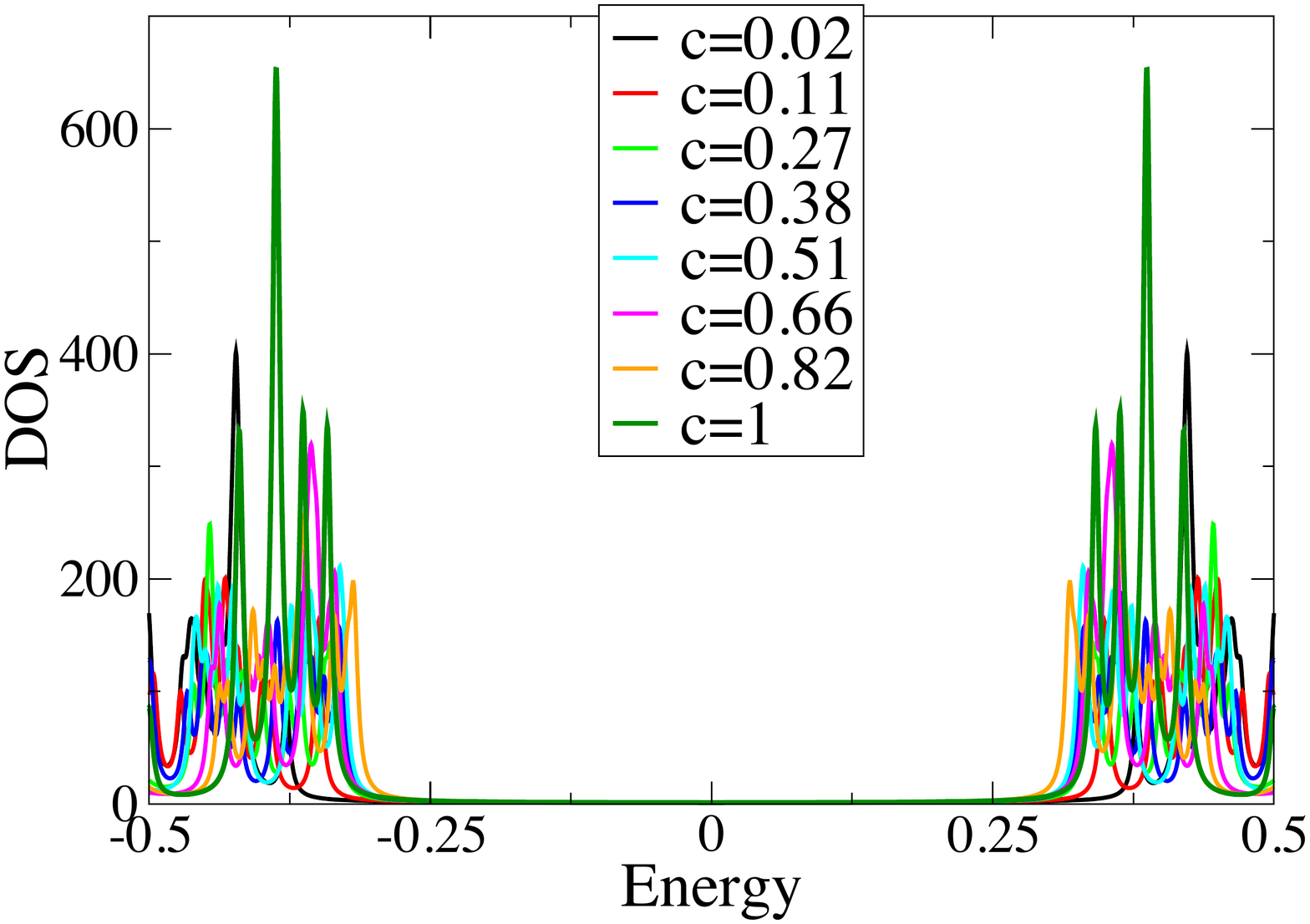}
\includegraphics[width=0.32\textwidth]{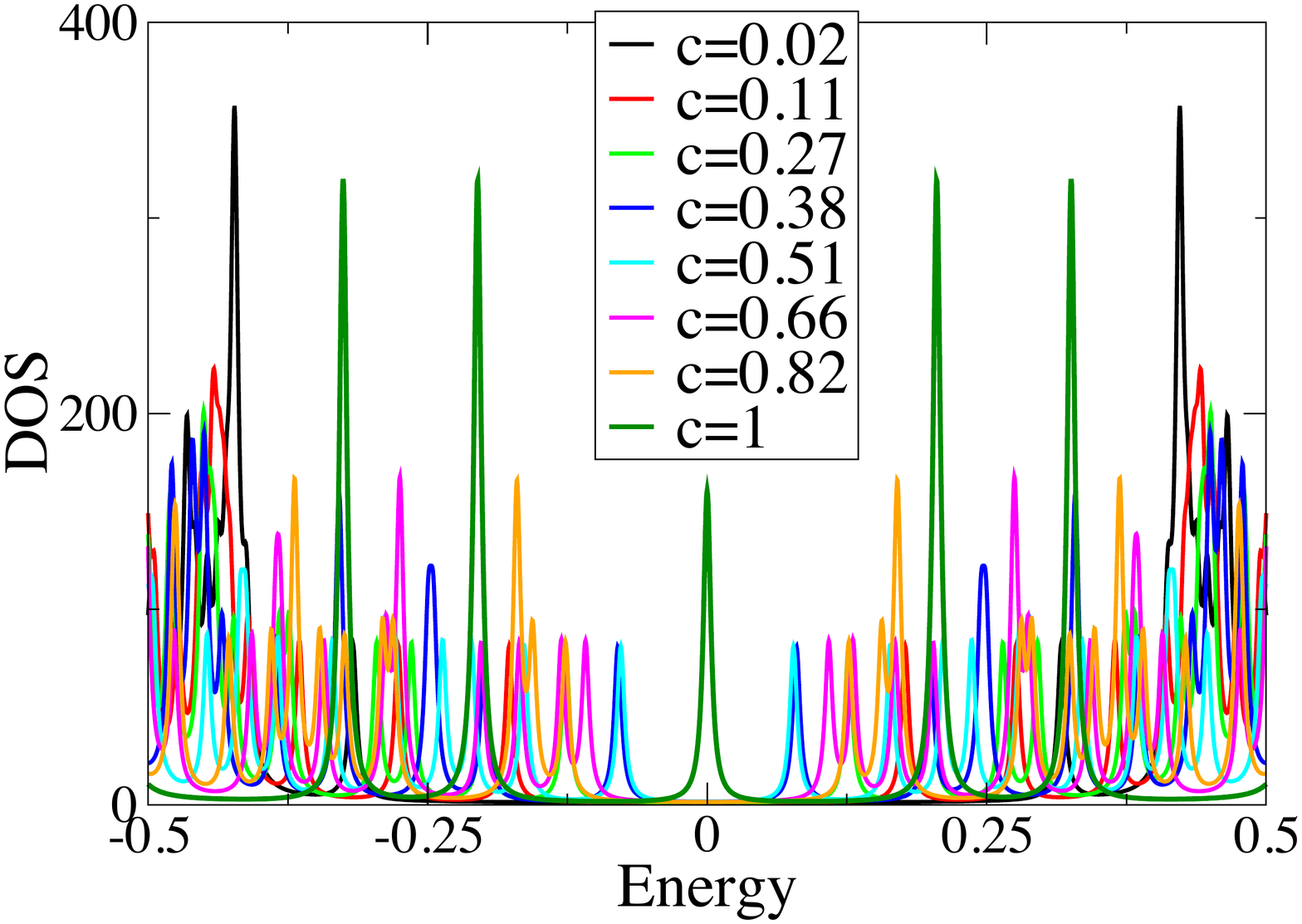}
\includegraphics[width=0.32\textwidth]{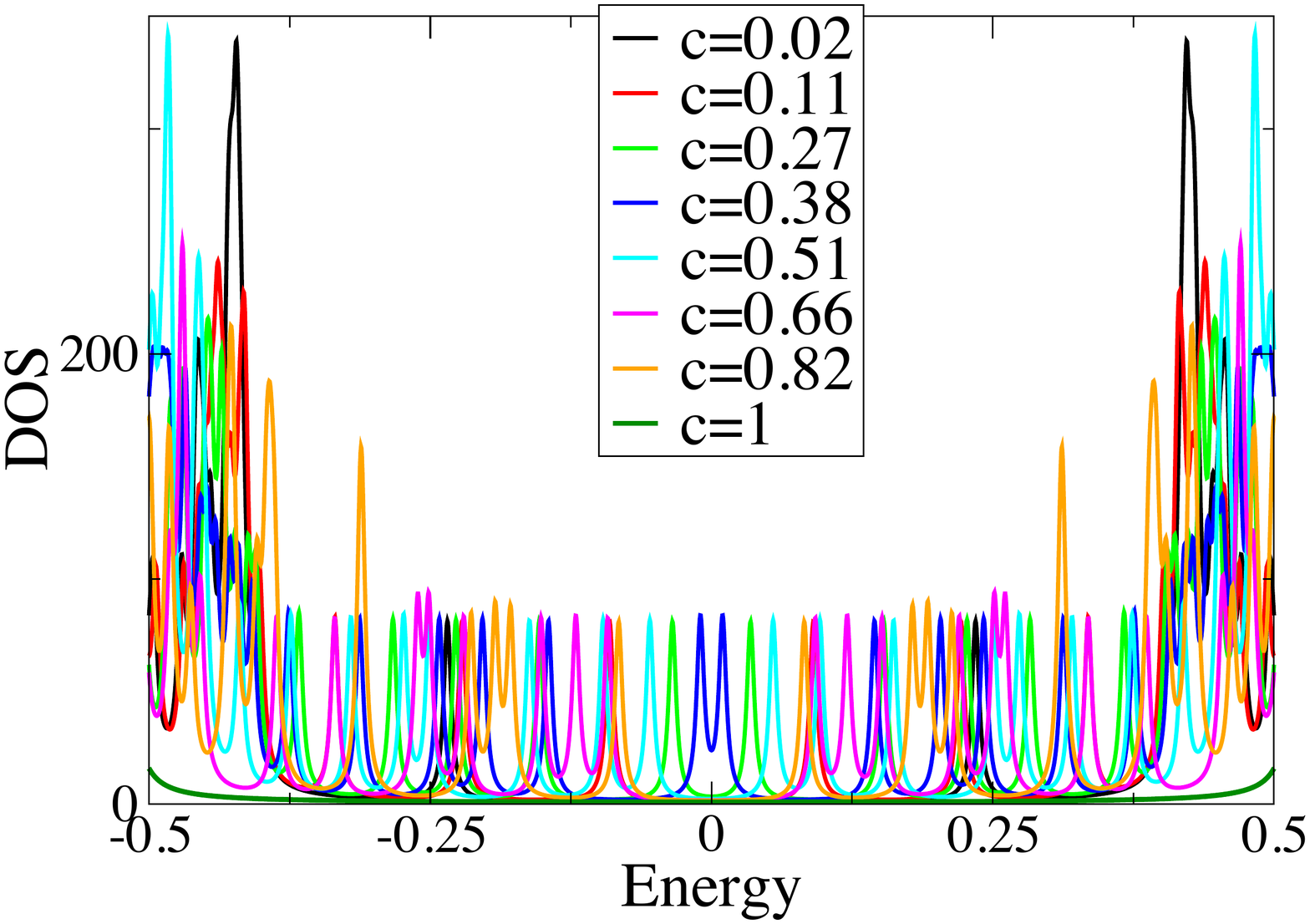}
\includegraphics[width=0.32\textwidth]{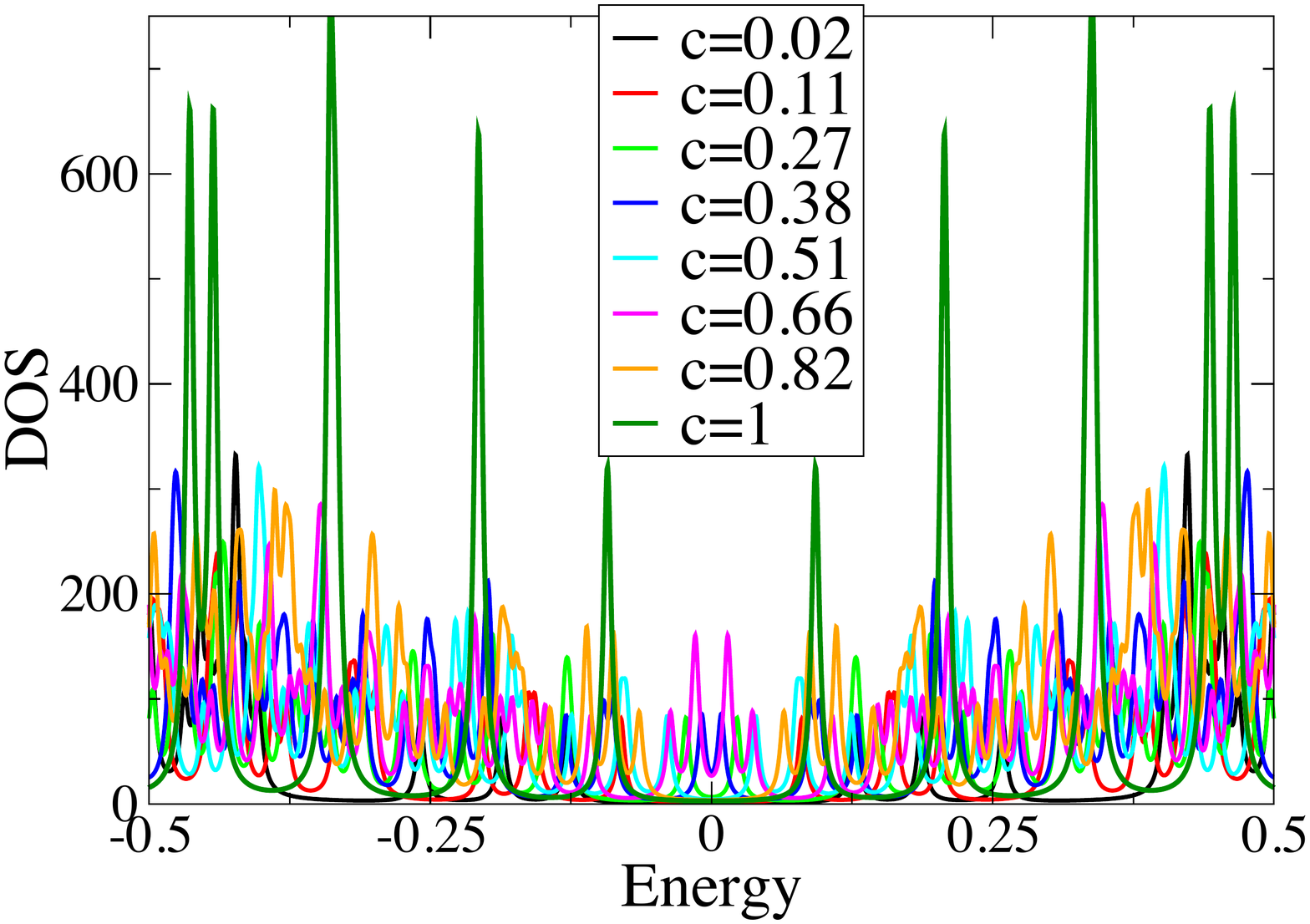}
\includegraphics[width=0.32\textwidth]{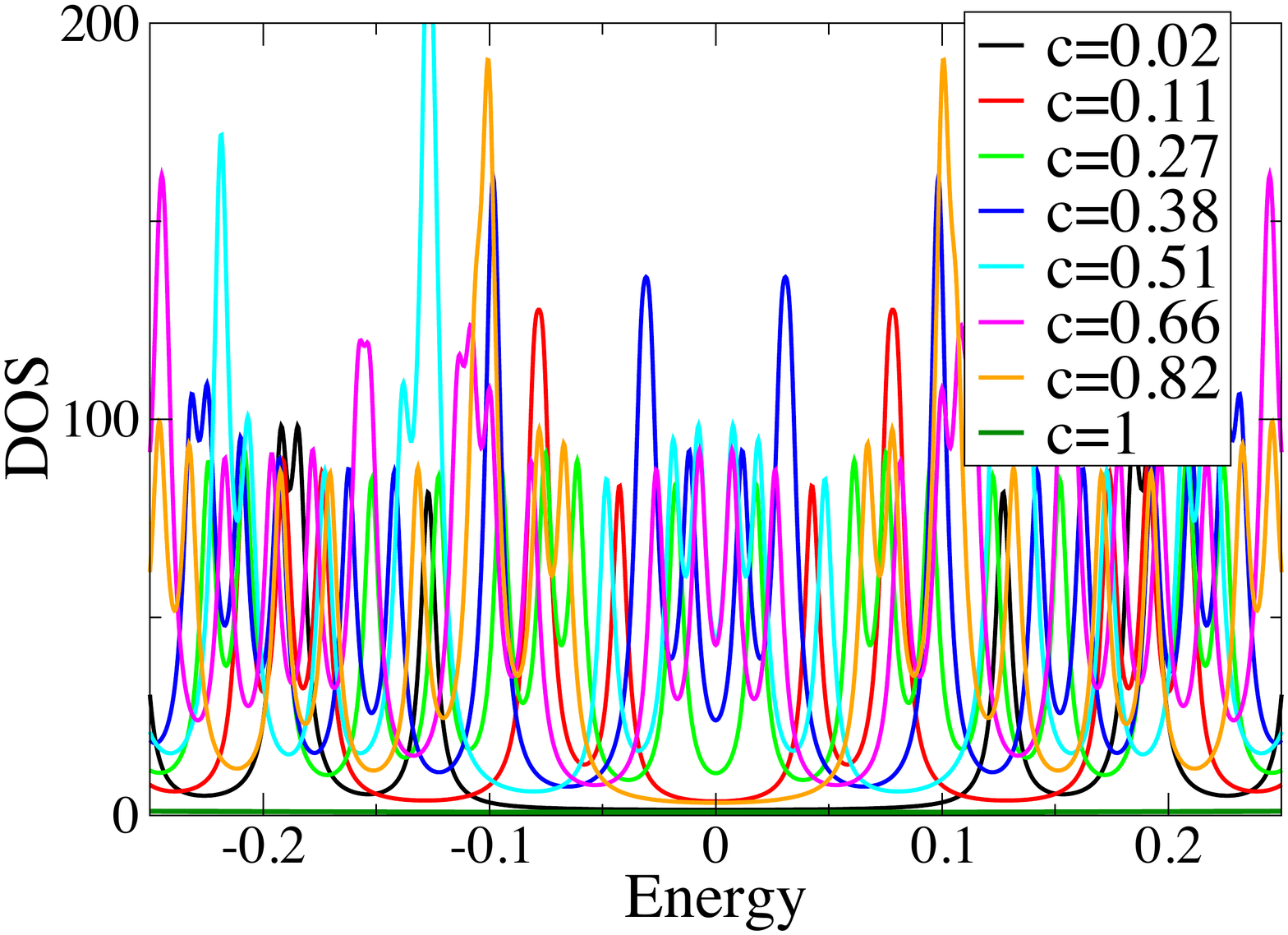}
\includegraphics[width=0.32\textwidth]{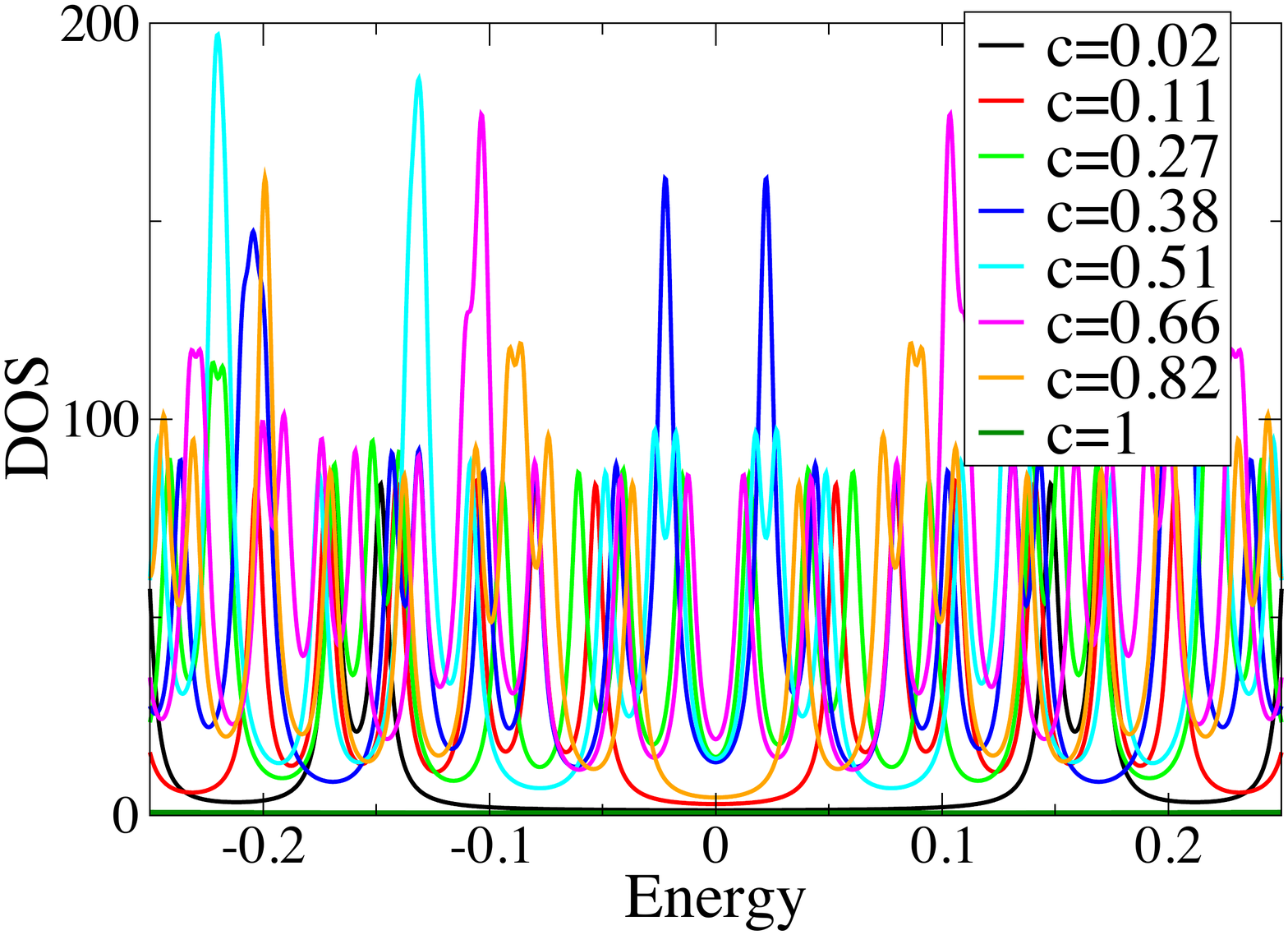}
\caption{\label{fig11}
(Color online) 
DOS for 
various magnetic fields $h_z=0.5,1,2,3,4,5$
and various concentrations 
$c_i=0.02,0.11,0.27,0.38,0.51,0.66,0.82,1$ for
$\mu=-3$. 
}
\end{figure*}

In Figs. \ref{fig9},\ref{fig10},\ref{fig11} we show results for the density of states (DOS)
for various cases. We see that often when a gap closes (or nearly closes) the energy levels approach
these low energies in pairs centered around zero energy. This splitting of the bands presumably
gives rise to Chern values that are half-integer. Something similar occurs in the calculation of
the Hall conductance of a vortex lattice at high magnetic fields in the Landau level description
calculated in the diagonal approximation, where as the gap amplitude approaches zero the limitting
value of the Hall conductance does not approach the normal state (integer) result, but a half-integer
value. This result is due to the doubling of the energy bands and associated filling of half the bands around the 
Fermi level \cite{hall}.
Something similar happens in graphene \cite{nuno}.

In Fig. \ref{fig9} we consider $\mu=0$, various concentrations and magnetic fields.
The case $c_i=1$ corresponds naturally to a uniform magnetic field. At $h_z=4$ the gap closes
due to the QPT between $C=-2$ and $C=0$ and there is a zero energy peak in the DOS.
Introducing a small concentration of magnetic impurities, states are introduced in the gap that
move to lower energies as the magnetic field at each impurity location increases. Increasing the
impurity concentration (the size of the magnetic island) the behavior is no longer monotonic. 
As the field increases, the subgap states move away from small energies and have 
values that exceed the gap at zero magnetic field. In the uniform case ($c_i=1$) the energy
gap is larger than the zero-field value.
The presence of low-energy
states at small concentrations leads to non-analyticities and a Chern number that is not well-defined,
as shown in Fig. \ref{fig7}. At larger concentrations the subgap density of states moves from small
energy to higher energies. The vicinity of $h_z=4$ is however close to the QPT at $c_i=1$, and
signals the transition between $C=-2$ and $C=0$, even at small and intermediate concentrations.
For $h_z>4$ the system is in a trivial phase independent of the impurity concentration,
as clearly illustrated in Fig. \ref{fig7}.

In Fig. \ref{fig10} we also consider $\mu=0$ but each panel has a fixed magnetic field and the impurity
concentration is varied. The case $h_z=4$ shows a filled gap for all $c_i$, as discussed above;
the same occurs for $h_z=5$, but, at least for high concentrations, the system is clearly in the trivial phase,
as shown by the value of $C=0$ of Fig. \ref{fig7}. Except at the transition points of the QPT, the gap
in the uniform case is large.
As $h_z$ increases the low-energy part of the gap gets increasingly filled but in a non-monotonic way as a 
function of the impurity concentration.
At the small value of $h_z=1$ we see that, excluding the lowest concentrations, increasing the size of
the magnetic island the low energy states increase in energy exceeding the value at zero field.
The same trend occurs for higher fields, but low energy states are found as one enters the
trivial phase $h_z \geq 4$. 

Finally we consider $\mu=-3$ in Fig. \ref{fig11}. The results for the Chern number in Fig. \ref{fig8} are
qualitatively the same, but now there are two QPT for $C=0$ to $C=1$, and to $C=-1$. At $c_i=1$ the
first transition occurs at $h_z=1$ and the second transition at $h_z=3$. The first transition is clearly
seen for $c_i=1$ with a zero-energy peak. However, the transition at $h_z=3$ is not seen. In the thermodynamic
limit the first transition occurs at $k_x=0$ while the second one at $k_x=\pi$ \cite{sato,tharnier}. The results shown
are obtained for a system of size $21 \times 21$. The first transition is captured by such a small system but the 
second one has the wrong parity. We have checked that a system of size $22 \times 22$ captures the second transition
(as it also does capture the first one) with a very small energy of $10^{-15}$.
Note the splitting of levels at low energies that explains the half-integer values for the Chern number.
At small fields ($h_z \leq 1$) the dependence on the impurity concentration is now the opposite of the case at half-filling:
a small concentration induces states that are of an energy that is larger than the zero-field case
and, as the concentration increases, the energies get smaller, but do not differ appreciably from the zero-field case.
Crossing the first QPT states appear in the gap down to very small energies, again in a non-monotonic way. 
At and after the QPT at 
$h_z=3$ (in the uniform case) there is a high density of states near zero energy that remains in the trivial phase.
This happens as long as the magnetic island does not fill the entire system. Close to $c_i=1$ the gap increases fast
and, in the uniform limit, it is large. 
Considering larger systems changes the details of the energy levels distributions but the qualitative analysis
remains the same.

\section{Summary}

The change of topology of a triplet superconductor due to the addition of magnetic
impurities was studied in this work. Under appropriate conditions the zero modes along the
borders of the superconductor are replaced by edge states localized inside the superconductor
or at the border and at the ends of a magnetic chain if long enough.
Increasing the density of impurities to magnetic islands of growing dimension leads
to chiral states that replace the helical states at zero magnetic field, as shown by
the non-vanishing Chern number.

The author acknowledges discussions with Eduardo Castro on the Chern number in real
space and support from FCT through grant UID/CTM/04540/2013.
Partial support from the Japan Society for the Promotion of Science is gratefully
acknowledged, as well as discussions with Norio Kawakami and Satoshi Fujimoto.

\end{document}